\documentclass[12pt]{article}

\newcommand{\VersionInformation}{}  
\InputIfFileExists{Debug}{}{}

\usepackage[utf8]{inputenc}
\usepackage{amsmath}
\usepackage{amsthm}
\usepackage{amsfonts}          
\usepackage{amssymb}           
\usepackage[mathscr]{eucal}
\usepackage{graphicx}
\usepackage{color}
\usepackage{hypbmsec}
\usepackage{fancyvrb}
\usepackage{sepnum}
\usepackage{xspace}
\usepackage{booktabs}
\usepackage{rotating}
\usepackage[isu,bf]{caption}
\newlength{\xtrawidth}
\newlength{\xtraheight}
\usepackage[all]{xy}           

\ifx\NoBackreferences\EMPTYMACRO
  \usepackage[backref,linktocpage,bookmarks]{hyperref}
\else
  \usepackage[linktocpage,bookmarks]{hyperref}
\fi

\ifx\DEBUG\EMPTYMACRO
\else
  \usepackage{showlabels}
\fi

\def\clap#1{\hbox to 0pt{\hss#1\hss}}

\def\mathrlap{\mathpalette\mathrlapinternal}

\def\mathrlapinternal#1#2{%
\rlap{$\mathsurround=0pt#1{#2}$}}

\makeatletter
  \def\adots{\mathinner{\mkern2mu\raise\p@\hbox{.}
      \mkern2mu\raise4\p@\hbox{.}\mkern1mu
      \raise7\p@\vbox{\kern7\p@\hbox{.}}\mkern1mu}}
\makeatother

\newcommand{\comma}[1]{\ensuremath{\sepnum{{.}}{{,}}{}{#1}}}

\newcommand{\eqdef}{%
  \mathrel{\lower.1mm
    \hbox{$\stackrel{\lower.424ex\hbox{\scriptsize def}}{=}$}}
}

\newcommand{\C}{\ensuremath{{\mathbb{C}}}}

\newcommand{\Z}{\mathbb{Z}}
\newcommand{\CP}{{\ensuremath{\mathop{\null {\mathbb{P}}}\nolimits}}}

\newcommand{\tors}{\ensuremath{\text{tors}}}

\DeclareMathOperator{\Span}{span}

\DeclareMathOperator{\rank}{rank}

\DeclareMathOperator{\Aut}{Aut}

\DeclareMathOperator{\img}{img}

\DeclareMathOperator{\Hom}{Hom}

\newcommand{\Osheaf}{\ensuremath{\mathscr{O}}}

\newcommand{\dP}[1]{\ensuremath{dP_{#1}}}

\DeclareMathOperator{\conv}{conv}
\DeclareMathOperator{\ord}{ord}

\newenvironment{descriptionlist}{%
\begin{list}%
{}%
{}}%
{\end{list}}

{\everymath{\displaystyle\everymath{}}\array}%
{\endarray}

\newtheorem{definition}{Definition}

\newcommand{\cone}[1]{\ensuremath{\left<#1\right>}}

\begin{document}
\begin{titlepage}
  \vspace*{-2cm}
  \VersionInformation
  \hfill
  \parbox[c]{5cm}{
    \begin{flushright}
    \end{flushright}
  }
  \vspace*{2cm}
  \begin{center}
    \Huge 
    Toric Elliptic Fibrations\\
    and F-Theory Compactifications
  \end{center}
  \vspace*{8mm}
  \begin{center}
    \begin{minipage}{\textwidth}
      \begin{center}
        \sc 
        Volker Braun
      \end{center}
      \begin{center}
        \textit{
          Dublin Institute for Advanced Studies\hphantom{${}^1$}\\
          10 Burlington Road\\
          Dublin 4, Ireland
        }
      \end{center}
      \begin{center}
        \texttt{Email: vbraun@stp.dias.ie}
      \end{center}
    \end{minipage}
  \end{center}
  \vspace*{\stretch1}
  \begin{abstract}
    The \comma{102581} flat toric elliptic fibrations over $\CP^2$ are
    identified among the Calabi-Yau hypersurfaces that arise from the
    \comma{473800776} reflexive 4-dimensional polytopes. In order to
    analyze their elliptic fibration structure, we describe the
    precise relation between the lattice polytope and the elliptic
    fibration. The fiber-divisor-graph is introduced as a way to
    visualize the embedding of the Kodaira fibers in the ambient toric
    fiber. In particular in the case of non-split discriminant
    components, this description is far more accurate than previous
    studies. The discriminant locus and Kodaira fibers of all
    \comma{102581} elliptic fibrations are computed. The maximal gauge
    group is $SU(27)$, which would naively be in contradiction with
    6-dimensional anomaly cancellation.
  \end{abstract}
  \vspace*{\stretch1}
\end{titlepage}
\tableofcontents
\listoffigures 	
\listoftables 	


\section{Introduction}
\label{sec:into}

F-theory~\cite{Vafa:1996xn, Morrison:1996na, Morrison:1996pp,
  Donagi:2009ra, Marsano:2011hv, Katz:2011qp} is a type of string
theory compactification, even though there is no fundamental
description available. However, there is a dictionary between the
low-energy gauge groups and the structure of elliptically-fibered
Calabi-Yau manifolds. For example, the ADE-classification of Kodaira
fibers corresponds to the ADE-gauge groups in a beautiful
correspondence. Further properties of the low-energy effective action
are encoded in higher-codimension degenerate fibers. Although known
for a long time, it has only recently been brought to the attention of
physicists that Kodaira's classification does not extend beyond
codimension-one degenerate fibers~\cite{MR690264, Esole:2011sm}.  In
fact, degenerate fibers in higher codimension have only been
classified under certain technical restrictions that are most likely
too restrictive for our purposes. One goal of this work is to present
a large number of examples of smooth elliptic fibrations and their
degeneration in various codimensions.

Likewise, our understanding of the consistent gauge theories is
incomplete. It has been suggested~\cite{Vafa:2005ui} that, in fact,
most gauge theories cannot be coupled to gravity in a consistent
manner. However, lacking any decisive criterion for which ones are and
are not consistent, it is difficult to make any decisive statement. In
order to say something definitive, one needs to restrict oneself to a
case where one has both strong restrictions on gauge theories as well
as reasonable control over the codimension-two and higher
degenerations of elliptic fibrations. In a beautiful
work~\cite{Kumar:2009ac, Kumar:2010am, Kumar:2010ru, Seiberg:2011dr},
it was pointed out that $6$-dimensional $N=1$ supergravities provide
such a setting: Three-dimensional elliptic fibrations are the first
dimension where codimension-two degenerations can occur, and
simultaneously there are very strong anomaly cancellation conditions
in the gauge theory. In particular, the simplest case of theories
without tensor multiplets~\cite{Kumar:2010am} is highly
constrained. Geometrically, this corresponds to elliptic fibrations
over $\CP^2$, which are likewise the most simple class of elliptic
threefolds. In this paper, we will try to address the geometric side
of these theories by classifying the hypersurfaces in toric varieties
that are elliptic fibrations over $\CP^2$.


\section{Toric Elliptic Fibrations}
\label{sec:EllFib}

\subsection{Toric Morphisms}
\label{sec:morphism}

The defining feature of a $d$-dimensional irreducible toric variety
$X_\Sigma$ is that it comes with a faithful algebraic torus action
\begin{equation}
  (\C^\times)^d \times X_\Sigma \to X_\Sigma
\end{equation}
such that there is a single maximal torus orbit $(\C^\times)^d \subset
X_\Sigma$. The combinatorics of how the finitely-many lower
dimensional orbits are glued to the boundaries of the maximal torus
orbit equals the combinatorial data of cones in a fan, and I will
frequently switch between torus orbits in $X_\Sigma$ and cones in the
fan $\Sigma$.

Having set the stage, let us now start by reviewing toric morphisms,
that is, \emph{toric} maps between toric varieties. These are maps
$\phi:X_{\Sigma_1}\to X_{\Sigma_2}$ between two irreducible toric
varieties that are both equivariant with respect to the torus action
and map the maximal torus of $X_{\Sigma_1}$ to the maximal torus
$X_{\Sigma_2}$. One can show~\cite{2000math.....10082H} that:
\begin{itemize}
\item Each fiber of a toric morphism is again a toric variety.
\item The fiber only depends on the torus orbit of the base point.
\item The generic fiber, that is, every fiber over the big torus orbit
  in the base, is irreducible and its embedding in the total space is
  again a toric morphism.
\item The degenerate fibers, that is, the fibers fixed by least one
  $(\C^\times)$-factor of the maximal torus of the base, are often
  reducible toric varieties.\footnote{Note that only irreducible toric
    varieties correspond to fans. A reducible toric variety is the
    result of gluing torus orbits of irreducible toric varieties by
    toric morphisms.}  Their embedding in the total space is not a
  toric morphism.
\end{itemize}
The data defining a toric morphisms is really the combinatorial
information of how the finitely many torus orbits map to each
other. This can be encoded in a morphism\footnote{By abuse of
  notation, we denote both maps by $\phi$ in the following.} $\phi:
\Sigma_1 \to \Sigma_2$ of fans, by which we mean a lattice map
$\Sigma_1 \subset N_1\to N_2 \supset \Sigma_2$ that maps cones into
cones, that is,
\begin{equation}
  \phi(\sigma_1) \subset \sigma_2
  \quad 
  \forall \sigma_1 \in \Sigma_1
  ,~
  \sigma_2 \in \Sigma_2
  .
\end{equation}
Toric geometry is a (covariant) functor from the category of fans and
fan morphisms to toric varieties and toric morphisms.

\subsection{Homogeneous Coordinates}

\begin{figure}
  \centering
  \includegraphics{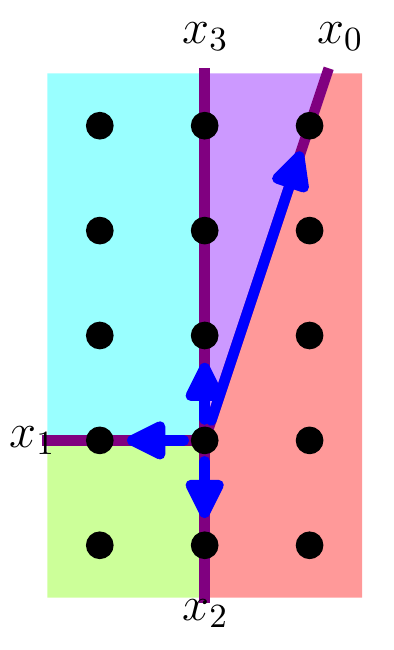}$~$\\
  $\vcenter{\hbox{\Huge $\downarrow$}}\;\varphi$\\
  \vspace{-10mm}
  \includegraphics{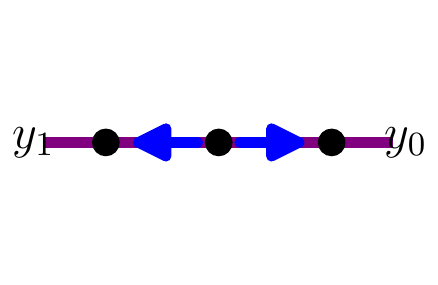}
  \vspace{-7mm}
  \caption{Toric fibration of the Hirzebruch surface $F_3$ over $\CP^1$.}
  \label{fig:Hirzebruch}
\end{figure}

A very convenient way of working with toric varieties are homogeneous
coordinates~\cite{Cox:1993fz}, which are generalizations of the usual
homogeneous coordinates on projective spaces (which happen to be
toric varieties). Roughly, for each ray spanning a one-dimensional
cone $\rho_i\in\Sigma(1)$ there exists a homogeneous coordinate
$z_i$. Certain subsets of the homogeneous coordinates are not allowed
to vanish simultaneously. Finally, we divide out a subgroup of
homogeneous rescalings to represent the toric variety as an algebraic
quotient
\begin{equation}
  X_\Sigma = 
  \frac{\C^{|\Sigma(1)|} -
    Z_\Sigma}{\Hom\big(A_{d-1}(X),\,\C^\times\big)}
  = 
  \frac{\C^{|\Sigma(1)|} -
    Z_\Sigma}{(\C^\times)^{\rank A_{d-1}(X)} \times A_{d-1}(X)_\tors}
  .
\end{equation}
Toric morphisms between smooth toric varieties can be written as
monomials in homogeneous coordinates. For example, take the Hirzebruch
surface $F_3$ fibered over $\CP^1$, see \autoref{fig:Hirzebruch}. Note
that there is a unique fan morphism. 
In terms of homogeneous coordinates, the base $\CP^1$ has the usual
homogeneous coordinates $[y_0:y_1]\in \CP^1$. The Hirzebruch surface
is given by
\begin{equation}
  F_3 =
  \big\{
  [x_0:x_1:x_2:x_3] 
  ~\big|~
  (x_0, x_1) \not=(0,0)
  ,~
  (x_2, x_3) \not=(0,0)
  \big\}
\end{equation}
subject to the homogeneous rescalings corresponding to the linear
relations between the generators. Let $\vec{x}_i$ be the primitive
lattice vector generating the ray corresponding to the homogeneous
coordinate $x_i$, then a basis for the linear relations is
\begin{equation}
  \vec{x}_0 + \vec{x}_1 + 3 \vec{x}_2 = 0
  ,\quad
  \vec{x}_2 + \vec{x}_3 = 0
  .
\end{equation}
The corresponding homogeneous rescalings are
\begin{equation}
  [x_0:x_1:x_2:x_3] =
  [\mu x_0:\mu x_1:\mu^3 x_2:x_3] =
  [x_0:x_1:\nu x_2:\nu x_3] 
  \quad
  \forall \mu,\nu \in \C^\times
  .
\end{equation}
To express the toric morphism $\varphi$ in terms of the homogeneous
coordinates, one needs to write the images of ray generators as
non-negative linear combinations of the base ray
generators. In~\autoref{fig:Hirzebruch}, this is
\begin{equation}
  \varphi(\vec{x}_i) = \sum_j \varphi_{ij} \vec{y}_j
  ,\quad
  (\varphi_{ij}) =
  \left(
    \begin{smallmatrix}
      1 & 0 \\
      0 & 1 \\
      0 & 0 \\
      0 & 0 
    \end{smallmatrix}
  \right)
  a
\end{equation}
and the corresponding map of homogeneous coordinates is
\begin{equation}
  \varphi:
  F_3 \to \C^2/\Z_2:
  \quad
  [x_0:x_1:x_2:x_3] \mapsto 
  \left[
    \prod_i x_i^{\varphi_{i0}} : \prod_i x_i^{\varphi_{i1}}
  \right]
  =
  [x_0:x_1]
\end{equation}

A point of the maximal torus orbit is characterized by all homogeneous
coordinates being non-zero. Moving fibers around by the torus-action
if necessary, we can the take all homogeneous coordinates to be
unity. Hence, a generic toric fiber is
\begin{equation}
  \varphi^{-1}([1:1]) = 
  \big\{ [1:1:x_2:x_3] ~\big|~  x_2, x_3 \in \C,~ 
  (x_2,x_3)\not=(0,0) \big\}
  = \CP^1
\end{equation}
Combinatorially, the generic fiber is given by the kernel fan of the
toric morphism $\varphi$, that is, by the set of all cones that map to
zero. In this example, the kernel fan consists of the two one-cones
corresponding to $x_2$, $x_3$, and the trivial cone. There are two
non-generic fiber, namely the fibers over $[y_0:y_1]=[1:0]$ and
$[0:1]$. They are
\begin{equation}
  \begin{split}
    \varphi^{-1}([1:0]) &= 
    \big\{ [1:0:x_2:x_3] ~\big|~  x_2, x_3 \in \C,~ 
    (x_2,x_3)\not=(0,0) \big\}
    = \CP^1
    ,\\
    \varphi^{-1}([0:1]) &= 
    \big\{ [0:1:x_2:x_3] ~\big|~  x_2, x_3 \in \C,~ 
    (x_2,x_3)\not=(0,0) \big\}
    = \CP^1
    .
  \end{split}
\end{equation}
Their embedding in $F_3$ is not a toric morphism, because the image is
not contained in the maximal torus of $F_3$. Due to the simplicity of
the example, the fibers over lower-dimensional torus orbits happen to
be again irreducible and, in fact, isomorphic to the generic
fiber. This means that the Hirzebruch surface is not only a
$\CP^1$-fibration over $\CP^1$, but, in fact, a $\CP^1$-bundle.

\begin{figure}[htbp]
  \centering
  \includegraphics{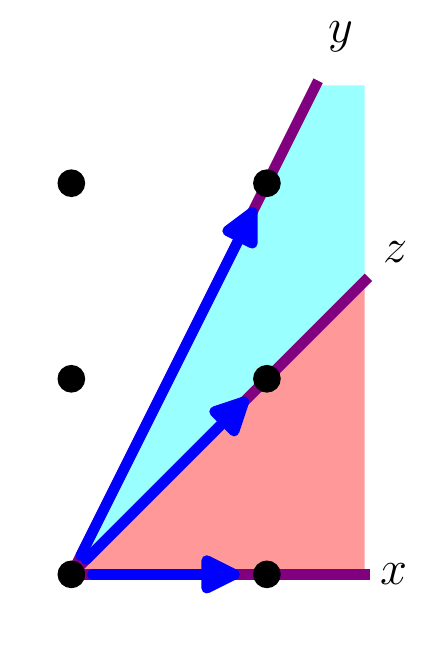}
  \hspace{1cm}
  \raisebox{3cm}{$\stackrel{\hbox{\Huge $\rightarrow$}}{\varphi}$}
  \hspace{1cm}
  \includegraphics{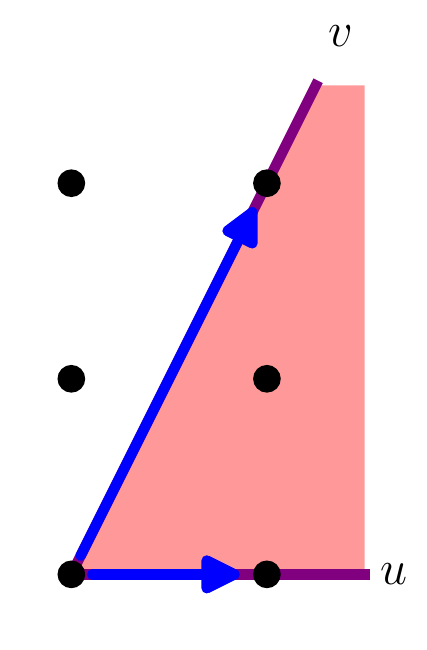}
  \caption{Blowup of $\C^2/\Z_2$.}
  \label{fig:ResolC2Z2}
\end{figure}

Another well-known example of a toric morphism is the blow-up of
\autoref{fig:ResolC2Z2}, which is the surjection
$\Osheaf_{\CP^1}(-2)\to \C^2/\Z_2$. The corresponding fan morphism is
depicted in \autoref{fig:ResolC2Z2}.
Expressing the image ray generators by the ray generators of the
image, one finds
\begin{equation}
  \varphi
  \left(
    \begin{matrix}
      \vec{x} \\ \vec{y} \\ \vec{z}
    \end{matrix}
  \right)
  =
  \left(
    \begin{matrix}
      1 & 0 \\
      \tfrac{1}{2} & \tfrac{1}{2} \\
      0 & 1 \\
    \end{matrix}
  \right)
  \left(
    \begin{matrix}
      \vec{u} \\ \vec{v}
    \end{matrix}
  \right)
  .
\end{equation}
Hence, the map can be written in terms of homogeneous coordinates as
\begin{equation}
  \label{eq:toricmorphismroot}
  \Osheaf_{\CP^1}(-2)\to \C^2/\Z_2:
  \quad
  [x:y:z] 
  ~\mapsto~
  \left[ 
    x \sqrt{z} : y \sqrt{z}
  \right]
  =
  [u:v]
  .
\end{equation}
Note that the map apparently involves a choice of square root, however
both signs lead to the same map since $[u:v]=[-u:-v]$ in $\C^2/\Z_2$.

There are $4$ torus orbits in $\img(\varphi)=\C^2/\Z_2$, corresponding
to the $4$ cones of the fan. The generic fiber is
\begin{equation}
  \varphi^{-1}([1:1]) = 
  \big\{ [1:1:1] \big\}
  ,
\end{equation}
the fibers over the two one-dimensional torus orbits $v=0$ and $u=0$
are 
\begin{equation}
  \varphi^{-1}([1:0]) = 
  \big\{ [1:1:0] \big\}
  ,\quad
  \varphi^{-1}([0:1]) = 
  \big\{ [0:1:1] \big\}
  ,
\end{equation}
and the fiber over the torus fixed point $u=v=0$ is 
\begin{equation}
  \label{eq:C2Z2fib00}
  \varphi^{-1}([0:0]) = 
  \big\{ [x:0:y] ~\big|~ x,y \in \C,~ 
  (x,y)\not=(0,0) \big\} =
  \CP^1
\end{equation}

\subsection{Fibrations of Polytopes}

A particularly useful class of toric varieties are the Gorenstein Fano
toric varieties. This means that they are both not too wildly singular
and have enough sections of the anticanonical bundle, such that a
anticanonical hypersurface is smooth after resolving the ambient space
singularities. They are the face fans of reflexive lattice polytopes,
or subdivisions of the face fan such that all additional rays are
generated by integral points of the polytope. The duality of reflexive
polytopes is mirror symmetry for the Calabi-Yau hypersurfaces.

Because the embedding of the generic fiber in the total space of a
toric fibration is again a toric morphism, the fibration can already
be seen on the level of the lattice polytope. Namely, the preimage of
the origin in the base fan is a lattice plane in the total space
polytope that intersects the reflexive polytope in a lattice
sub-polytope containing the origin as a relative interior point. Note
that there are only finitely many lattice sub-polytopes since each
vertex must be one of the finitely many integral points of the total
space polytope. Hence, it is a finite combinatorial problem to
enumerate all lattice sub-polytopes in a lattice polytope. The
embedding of the lattice sub-polytope is the part of the toric data
that is visible just on the level of polytopes, without specifying the
details of the triangulation. In the following, we refer to this as a
\emph{fibration of polytopes}. However, note that there is no notion
of a base of the fibration when talking about polytopes alone. Indeed,
as we saw in the toric morphism \autoref{fig:ResolC2Z2}, the rays of
the domain fan need to map to rays of the codomain fan. In particular,
this means that the integral points of the total space polytope need
not map to integral points of any base polytope.

Note that it is important to identify fibrations that only differ by a
lattice automorphism in order to not overcount the number of
fibrations. For example, take the 24-cell, which is the reflexive
4-dimensional polytope with the largest symmetry
group~\cite{KreuzerSkarkeReflexive, Braun:2011hd}. Naively, the
24-cell lattice polytope has 34 fibrations with two-dimensional
fibers. They divide into 18 fibrations whose fiber is a lattice square
(the lattice polygon defining $\CP^1\times\CP^1$) and 16 fibrations
whose fiber is a lattice hexagon (defining $\dP{6}$, the del Pezzo
surface obtained by blowing up $\CP^2$ at 3 points). However, note the
lattice symmetry group of the 24-cell is the Weyl group of $F_4$,
which has order $1152$. By definition, the automorphism group fixes
the 24-cell, but generally maps sub-polytopes to other
sub-polytopes. Identifying the orbits of the fibrations, one finds
that there are indeed only two different fibrations: One whose fiber
is a square, and one whose fiber is a hexagon. In the following, we
will always count the number of fibrations modulo automorphisms.

The naive algorithm to enumerate all $d$-dimensional fibers is to
iterate over all linearly independent $d$-tuples of lattice points of
the total space. They define a lattice $d$-plane. Now compute the
intersection of the $d$-plane with the ambient polytope; If all
vertices are integral then it defines a fibration. An important
optimization over the naive algorithm is to note that one can take the
$d$ vertices of the fiber to lie all on the same facet of the
fiber. Hence, it suffices to iterate over $d$-tuples that
simultaneously saturate one of the ambient inequalities.

It is computationally feasible to enumerate all fibrations of the
\comma{473800776} reflexive 4-dimensional polytopes. There are
approximately an order of magnitude more fibrations than polytopes,
though we cannot offer a precise number since we have not modded out
the automorphisms for all of them. PALP~\cite{Kreuzer:2002uu} has an
option to enumerate fibrations, but since the author does not
understand some of the output the algorithm was implemented in
Sage~\cite{Sage, BraunHampton:polyhedra}. See \autoref{sec:flat} for
additional restrictions that were placed on the fibrations for the
purposes of this paper, and for the results of the search.

\subsection{Torus Fibrations}
\label{sec:TorusFib}

By a torus fibration we will always denote a fibration whose generic
fiber is a real torus $T^2 = \C/(\Z+\tau \Z)$. Since $T^2$ is not a
toric variety, this cannot be realized by the fibers of a toric
morphism. This is completely analogous to the fact that a toric
variety itself is never a Calabi-Yau manifold, which is why one has to
study hypersurfaces or complete intersections in toric varieties
(which can be Calabi-Yau manifolds).

Therefore, in the following we will consider the situation where
\begin{itemize}
\item $\pi: X_\Sigma\to B$ is a toric morphism with, generically,
  complex 2-dimensional fibers $\pi^{-1}(b)$, $b\in B$.
\item $Y\subset X_\Sigma$ is a Calabi-Yau fourfold hypersurface or,
  more generally, complete intersection.\footnote{However, for the
    purposes of this paper we restrict ourselves to hypersurfaces.}
\item $Y\cap \pi^{-1}(b) \simeq T^2$ is a real torus (with an induced
  complex structure, of course) for a generic point $b\in B$.
\end{itemize}


\section{Flat Fibrations}
\label{sec:flat}

\subsection{Kodaira vs.\ Miranda}

Kodaira~\cite{KodairaII, KodairaIII} determined the structure of
elliptically fibered surfaces by classifying the potential degenerate
fibers in codimension one, which follow an ADE-pattern. If one wants
to investigate compactifications of F-theory to six dimensions, that
is, on an elliptically fibered threefold, then the degenerate fibers
sit over the discriminant curve in the base. At a generic point of the
curve, one can simply pick a transverse direction and reduce the local
structure back to Kodaira's case. But the curve is
almost\footnote{Sometimes it is claimed that the discriminant curve is
  always singular, or that it always contains an $I_1$ component. The
  covering space of the $\Z_3\times\Z_3$ manifold~\cite{dP9Z3Z3,
    Braun:2007vy, Braun:2007xh, Braun:2007tp} is a counterexample to
  both of those claims.} always singular, so there are codimension-two
loci in the base where Kodaira's classification is not applicable. In
fact, there is no classification of codimension-two degenerate fibers
in general. However, under special circumstances there is. In
particular, there is a classification of codimension-two degenerate
fibers~\cite{MR690264} under the provision that the elliptic fibration
is flat, that the discriminant has only normal crossings, and that the
$j$-invariant of the elliptic fibration is well-defined. Even with all
these restrictions, there is an infinite family of non-Kodaira
degenerate fibers.

So far, I only mentioned the local structure of elliptic
fibrations. The Miranda models of the degenerate fibers tell us,
starting from the (singular) Weierstrass model, how the degenerate
fibers in the resolved manifold look like. We are, of course,
interested in compact threefolds. In order to classify the
elliptically-fibered Calabi-Yau threefolds, one would then first have
to classify all Weierstrass models with allowable singularities in the
discriminant such that the Weierstrass model can be resolved into a
smooth elliptically-fibered Calabi-Yau threefold, similar to was done
in~\cite{Morrison:2011mb} for $SU(n)$ gauge groups.

For the purposes of this paper, I will be going the opposite route and
start with smooth elliptically fibered Calabi-Yau threefolds. By far
the largest class of such manifolds are the toric
hypersurfaces~\cite{KreuzerSkarkeReflexive}, and I will focus on them
in the following.

\subsection{Toric Fibrations and Polytopes}

Restricting oneself to flat fibrations, that is, fibrations whose
fiber dimension is constant, is very natural if one wants to
investigate fibrations over a particular base. Otherwise, one could
always compose the fibration $X\to B$ with a blow-down $\pi:B\to
\hat{B}$ to get a fibration $X\to\hat{B}$. So, in particular, any
fibration over a blow-up of $\CP^2$ gives rise to a fibration over
$\CP^2$. However, if $X\to B$ was flat then the induced fibration
$X\to\hat{B}$ is most certainly not: The dimension of the fiber over
the blown-up point $\hat{b}\in \hat{B}$ jumps from $\dim(X)-\dim(B)$
to $\dim(X)-\dim(B)+\dim(\pi^{-1}(\hat{B}))$. In other words, to study
fibrations over a particular base (here: $\CP^2$), one should divide
up the fibrations into fibrations that are flat\footnote{Or, at least,
  cannot be flattened any further by blowing up the base.} on $\CP^2$,
$\CP^2$ blown up at one point, $\CP^2$ blown up at two points,
$\dots$. For the purposes of this paper, I will restrict therefore to
flat fibrations over $\CP^2$, and leave the more complicated cases for
future work.

In terms of toric geometry, we have already encountered the blowup
$\Osheaf_{\CP^1}(-2)\to\C^2/\Z_2$ an example of a non-flat fibration,
see \autoref{fig:ResolC2Z2}. The reason for why the fiber dimension is
not constant in this example is that one of the rays of the domain fan
maps to a higher-dimensional cone (in this case, the $2$-cone
$\cone{u,v}$) of the codomain fan. This means that there is a point in
the base (the torus orbit corresponding to the $2$-cone) whose fiber
is given by the vanishing of a single homogeneous coordinate, see
eq.~\eqref{eq:C2Z2fib00}. Clearly, this cannot be a flat fibration. A
necessary criterion for a flat fibration is that the rays of the
domain fan map either to zero or the rays of the codomain fan, but not
into any higher-dimensional cone. A necessary and sufficient
criterion~\cite{2000math.....10082H} is that every primitive cone of
the domain fan (not just the one-dimensional ones) maps bijectively to
its image cone.

Therefore, for \emph{flat} fibrations we can read of the base rays
from the polytope alone, without having to triangulate the total space
polytope: The rays of the base fan must be the images of the rays of
the total space fan.

\subsection{Classification of Fibered Polytopes}

As we saw above, for a flat fibration the rays of the base fan are
determined by the rays of the total space fan. For the purposes of
this paper, we will be interested in the Gorenstein Fano
$4$-dimensional toric varieties fibered over $\CP^2$. As with all
toric surfaces, the whole fan of $\CP^2$ is determined by the
rays. Furthermore, we want to have a smooth Calabi-Yau
hypersurface. For this, we need to subdivide the face fan of the
reflexive 4-dimensional polytope such that all integral points that
are not interior to facets\footnote{A one-dimensional cone generated
  by a point in the interior of a facet corresponds to a toric divisor
  that does not intersect the Calabi-Yau hypersurface, so it can be
  blown-down without inducing a singularity on the hypersurface.} span
a ray.

To summarize, on the level of polytopes we can enumerate the flat
fibrations over $\CP^2$ by the following steps. For all reflexive
4-dimensional polytopes $P$:
\begin{itemize}
\item Find all lattice sub-polytopes $S\subset P$
\item Project all integral points not interior to a facet of $P$.
\item Test whether the projected points span the rays of the fan of
  $\CP^2$.
\item Identify fibrations that map to each other by the action of
  $\Aut_\Z(P) = \Aut(P)\cap GL(4,\Z)$.
\end{itemize}
Searching this way through the list of \comma{473800776} reflexive
4-dimensional polytopes, we find \comma{102581} distinct fibered
polytopes corresponding to flat fibrations over $\CP^2$.

\begin{figure}
  \centering
  \vspace{-1cm}
  \begin{sideways}
    \includegraphics[width=21cm]{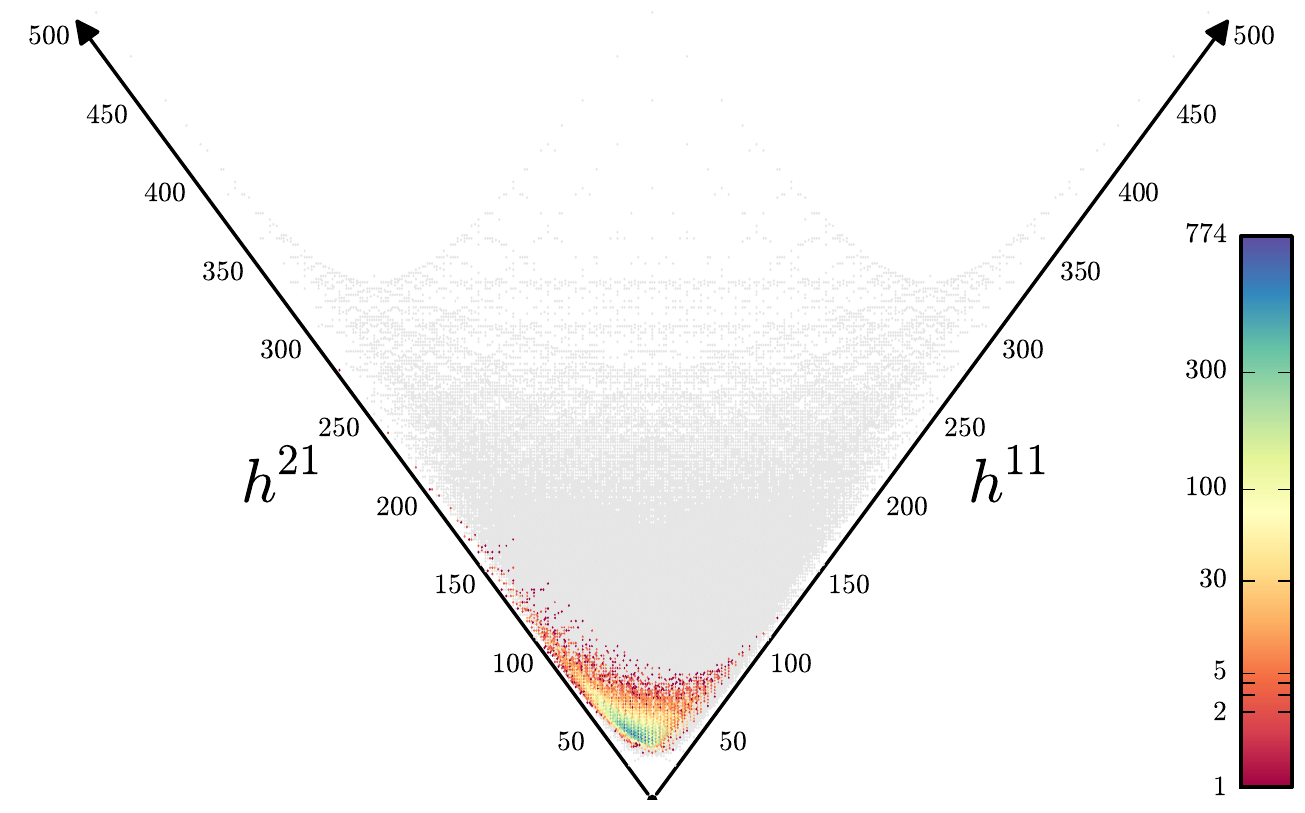}
  \end{sideways}
  \caption{The distribution of flat toric elliptic fibrations with
    base $\CP^2$.}
  \label{fig:bigplot}
\end{figure}

\begin{figure}
  \centering
  \vspace{-1cm}
  \begin{sideways}
    \includegraphics[width=21cm]{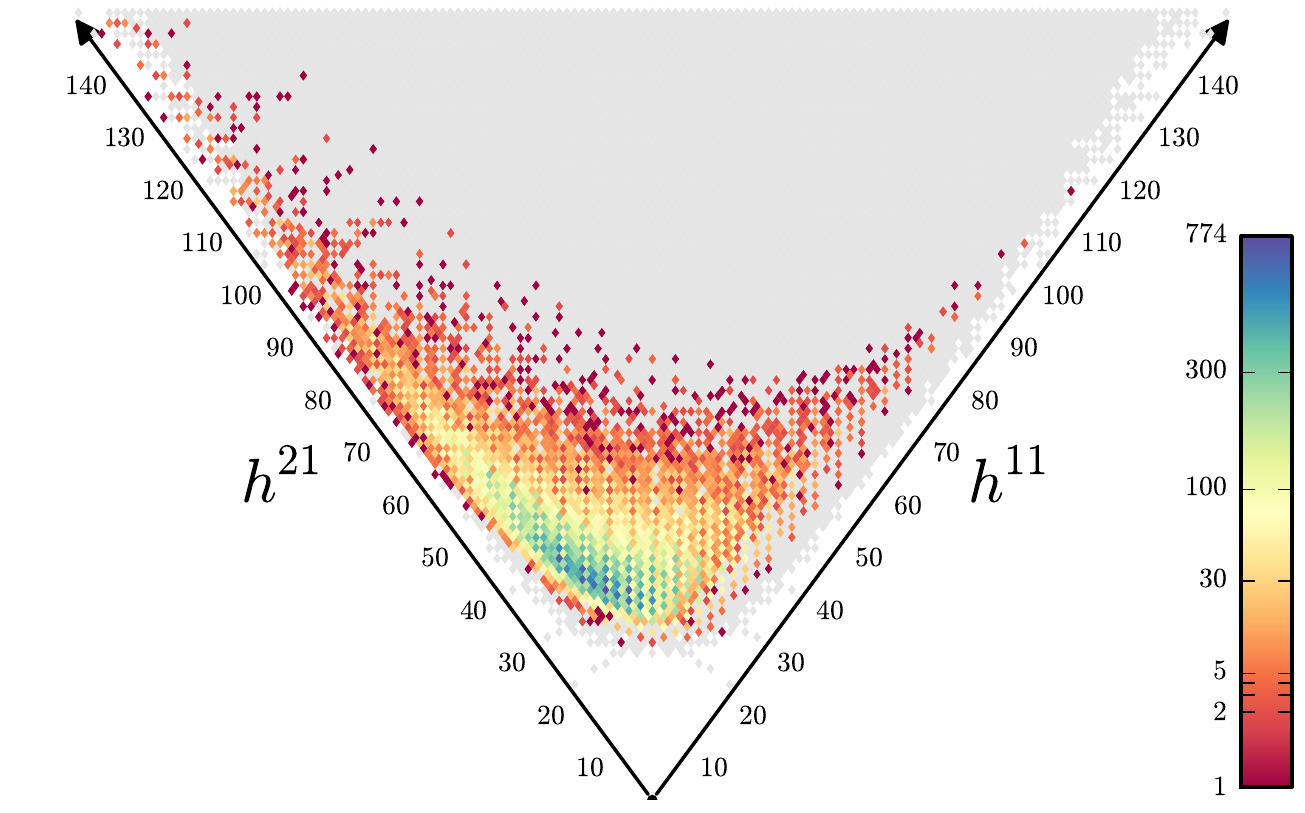}
  \end{sideways}
  \caption{The region of heights $h^{11}+h^{21}\leq 150$ in
    \autoref{fig:bigplot}.}
  \label{fig:zoomplot}
\end{figure}

The largest number of distinct fibered polytopes ($774$) is found for
the Hodge numbers $h^{11}=14$, $h^{21}=26$. The distribution of Hodge
numbers is shown in Figures~\ref{fig:bigplot} and~\ref{fig:zoomplot}.

\subsection{Weierstrass Models}

Before passing to explicit examples where the complete geometry will
be specified, there is one more piece of information that does not
depend on the details of how the fibration of polytopes is resolved
into a fibration of toric varieties. This is the Weierstrass model of
the elliptic fibration, obtained by bringing the hypersurface equation
into Weierstrass form $y^2=x^3+ax+b$ over the maximal torus of the
base. Obtaining the correct Weierstrass form depends on having enough
rays in the fan of the toric variety, but is otherwise independent of
the details of the details of the fan.

In terms of homogeneous coordinates, it is convenient to use
projective coordinates $[u:v:w]\in\CP^2$ for the base $\CP^2$ and
affine coordinates $(x,y)$ on the fiber. Then pick a parametrization
of the maximal torus of the total space fan such that
\begin{itemize}
\item $\vec{u}$, $\vec{v}$, and $\vec{w}$ map to the three generators
  of the base $\CP^2$-fan.
\item If the fiber fan is the fan of $\CP^2$, the $2$-cone
  $\cone{x,y}$ can be any $2$-cone. 
\item If the fiber fan is a blow-up of $\CP^2$, pick $\cone{x}$ and
  $\cone{y}$ to be cones that survive after blowing down to $\CP^2$.
\item Otherwise, for example if the fiber fan is $\CP^1\times\CP^1$,
  pick suitable coordinates to bring the (not necessarily cubic)
  equation into Weierstrass form, see \autoref{sec:Weierstrass} for
  how this can be done for any fiber reflexive polygon.
\end{itemize}
Having chosen $3+2$ rays in this manner, we just need to set all other
homogeneous coordinates equal to one in the hypersurface equation. The
result is a cubic in $x$, $y$ that can easily be brought into
Weierstrass form. In the remainder of this paper, we will now look at
three increasingly more complicated examples of how toric elliptic
fibrations can be analyzed.


\section{An Example of a Toric Elliptic Fibration}
\label{sec:example}

\subsection{Fibration of the Polytope}
\label{sec:simple}

\begin{figure}
  \centering
  \includegraphics{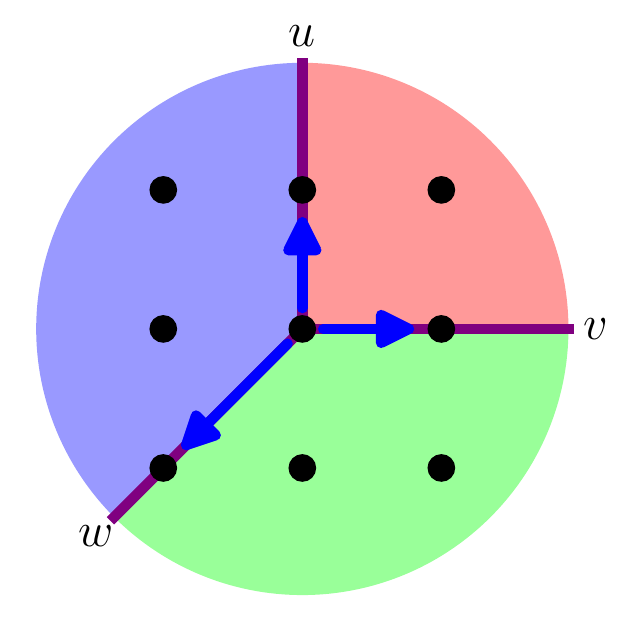}
  \caption{The fan of $\CP^2$.}
  \label{fig:P2}
\end{figure}

\begin{figure}
  \centering
  \includegraphics{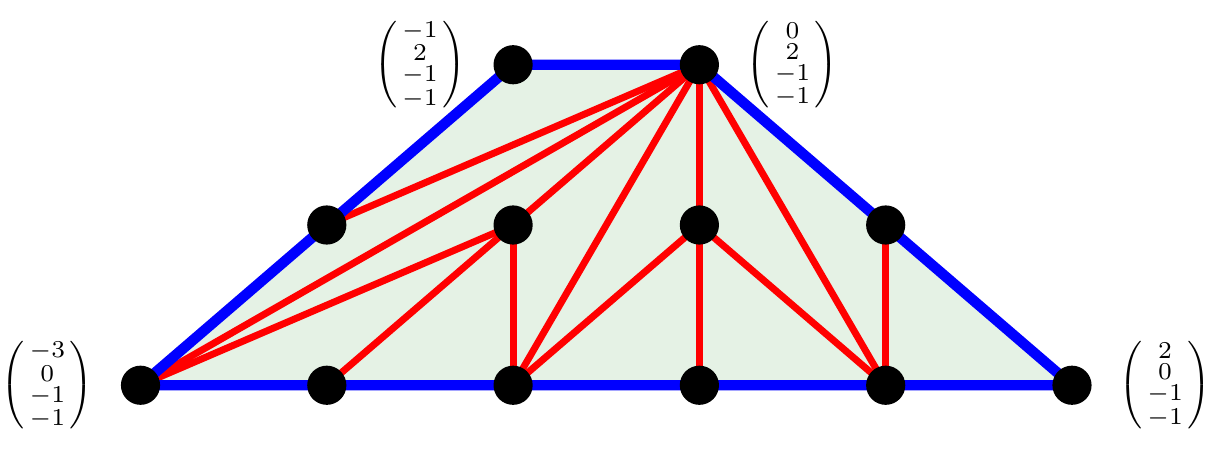}
  \caption{The two-face $F\subset P$ and its triangulation used in
    \autoref{sec:resolv}.}
  \label{fig:Ftriang}
\end{figure}

As the first example, consider the reflexive polytope with vertices
\begin{equation}
  P = \conv\left\{
  \left(\begin{smallmatrix}
      -3 \\     0 \\      -1 \\      -1
    \end{smallmatrix}\right),
  \left(\begin{smallmatrix}
      -1 \\      2 \\      -1 \\      -1
    \end{smallmatrix}\right),
  \left(\begin{smallmatrix}
      0 \\      -1 \\      0 \\      0
    \end{smallmatrix}\right),
  \left(\begin{smallmatrix}
      0 \\      0 \\      0 \\      1
    \end{smallmatrix}\right),
  \left(\begin{smallmatrix}
      0 \\      0 \\      1 \\      0
    \end{smallmatrix}\right),
  \left(\begin{smallmatrix}
      0 \\      1 \\      0 \\      0
    \end{smallmatrix}\right),
  \left(\begin{smallmatrix}
      0 \\      2 \\      -1 \\      -1
    \end{smallmatrix}\right),
  \left(\begin{smallmatrix}
      1 \\      0 \\      0 \\      0
    \end{smallmatrix}\right),
  \left(\begin{smallmatrix}
      2 \\      0 \\      -1 \\      -1
    \end{smallmatrix}\right)
  \right\}.
\end{equation}
In addition to the vertices and the origin, the lattice polytope $P$
has 9 further integral points
\begin{equation}
  P \owns 
  \left(\begin{smallmatrix}
      -2 \\0 \\-1 \\-1
    \end{smallmatrix}\right),
  \left(\begin{smallmatrix}
      -2 \\1 \\-1 \\-1
    \end{smallmatrix}\right),
  \left(\begin{smallmatrix}
      -1 \\0 \\-1 \\-1
    \end{smallmatrix}\right),
  \left(\begin{smallmatrix}
      -1 \\0 \\0 \\0
    \end{smallmatrix}\right),
  \left(\begin{smallmatrix}
      -1 \\1 \\-1 \\-1
    \end{smallmatrix}\right),
  \left(\begin{smallmatrix}
      0 \\0 \\-1 \\-1
    \end{smallmatrix}\right),
  \left(\begin{smallmatrix}
      0 \\1 \\-1 \\-1
    \end{smallmatrix}\right),
  \left(\begin{smallmatrix}
      1 \\0 \\-1 \\-1
    \end{smallmatrix}\right),
  \left(\begin{smallmatrix}
      1 \\1 \\-1 \\-1
    \end{smallmatrix}\right)
  ,
\end{equation}
none of which are interior to a facet of $P$. Moreover, $P$ is a
lattice polytope fibration with respect to the sub-polytope
\begin{equation}
  P \cap \big(\Z\oplus\Z\oplus\{0\}\oplus\{0\}\big) = 
  \conv\left\{
    \left(\begin{smallmatrix}
        -1 \\      0 \\      0 \\      0
      \end{smallmatrix}\right),
    \left(\begin{smallmatrix}
        0 \\      -1 \\      0 \\      0
      \end{smallmatrix}\right),
    \left(\begin{smallmatrix}
        0 \\      1 \\      0 \\      0
      \end{smallmatrix}\right),
    \left(\begin{smallmatrix}
        1 \\      0 \\      0 \\      0
      \end{smallmatrix}\right)
  \right\}.
\end{equation}
which we recognize as the lattice polygon of $\CP^1\times\CP^1$. The
lattice projection onto the base is, clearly,
\begin{equation}
  \varphi = 
  \begin{pmatrix}
    0 & 0 & 1 & 0 \\
    0 & 0 & 0 & 1 \\
  \end{pmatrix}
  ,
\end{equation}
and all integral points of $P$ not interior to facets map to the
standard rays of the fan of $\CP^2$, whose rays we label $\cone{u}$,
$\cone{v}$, and $\cone{w}$ as in \autoref{fig:P2}. 
What makes $P$ a particularly simple example is that there is exactly
one point over the two base rays $\cone{u}$ and $\cone{v}$. Hence, the
fibers over the corresponding torus orbits $\{u=0,wv\not=0\}$ and
$\{v=0,uw\not=0\}$ in the base $\CP^2$ are the same as the generic
fiber. Only over the toric divisor $\{w=0\}\subset \CP^2$ do we get a
more interesting fiber. More specific, there are $19$ integral points
in $P$:
\begin{itemize}
\item one point is over $\vec{u}$ and $\vec{v}$ each,
\item the fiber polytope $\varphi^{-1}(0)$ consists of 5 points, the
  origin and the four vertices of the $\CP^1\times\CP^1$ lattice
  square,
\item the remaining $12$ integral points are contained in the $2$-face
  $F = P \cap \varphi^{-1}(\vec{w})$, see \autoref{fig:Ftriang}.
\end{itemize}
and much of the information about the triangulation of $P$, that is,
the subdivision of the face fan of $P$, is contained in the
triangulation of this two-face $F$.

The simplest toric variety one can construct from $P$ is its face fan
(14 generating cones), but this toric variety is not fibered. The
problem is that $\varphi$ projects some $2$-faces of $P$ to
$\conv\{\vec{u},\vec{v},\vec{w}\}$, so the corresponding cone of the
face fan is not contained in any single cone of the base. However,
there is a well-defined procedure to subdivide the face fan along the
half-planes $\varphi^{-1}(\cone{u})$, $\varphi^{-1}(\cone{v})$, and
$\varphi^{-1}(\cone{w})$ that will lead to the minimal fibered toric
variety, that is, the coarsest partial resolution of the face fan such
that the toric variety is fibered (18 generating cones in this
example).

\subsection{Weierstrass Model}
\label{sec:P1xP1Weier}

The minimal fibered toric variety and any contained anticanonical
hypersurface is still very singular, but it is good enough to
determine the Weierstrass model. Following \autoref{sec:Weierstrass},
let us parametrize the maximal torus by picking five homogeneous
coordinates
\begin{equation}
  \vec{u} = 
  \left(\begin{smallmatrix}
      0 \\      0 \\      0 \\      1
    \end{smallmatrix}\right),\quad
  \vec{v} =
  \left(\begin{smallmatrix}
      0 \\      0 \\      1 \\      0
    \end{smallmatrix}\right),\quad
  \vec{w} = 
  \left(\begin{smallmatrix}
      -1 \\      2 \\      -1 \\      -1
    \end{smallmatrix}\right),\quad
  \vec{x} = 
  \left(\begin{smallmatrix}
      1 \\      0 \\      0 \\      0
    \end{smallmatrix}\right),\quad
  \vec{y} = 
  \left(\begin{smallmatrix}
      0 \\      1 \\      0 \\      0
    \end{smallmatrix}\right).
\end{equation}
The dual polytope $P^*$ has $35$ integral points, hence the equation
of a Calabi-Yau hypersurface has $35$ distinct monomials in the
homogeneous coordinates. Setting all other homogeneous variables to
unity, the 
\begin{equation}
  \label{eq:monomials}
  \begin{gathered}
    \{
    u w^5 y^2,\ v w^5 y^2,\ w^6 y^2,\ u^3 w^2 x y^2,\ u^2 v w^2 x y^2,\ 
    u v^2 w^2 x y^2,\ v^3 w^2 x y^2,\ 
    u^2 w^3 x y^2,\ 
    \\
    u v w^3 x y^2,\ v^2 w^3 x y^2,\ u w^4 x y^2,\ 
    v w^4 x y^2,\ w^5 x y^2,\ 
    w^4 x^2 y^2,\ u w^3 y,\ 
    v w^3 y,\ 
    \\
    w^4 y,\ u^3 x y,\ u^2 v x y,\ 
    u v^2 x y,\ v^3 x y,\ 
    u^2 w x y,\ u v w x y,\ v^2 w x y,\ u w^2 x y,\ 
    \\
    v w^2 x y,\ 
    w^3 x y,\ w^2 x^2 y,\ u w,\
    v w,\ w^2,\ u x,\ v x,\ w x,\ x^2
    \}
  \end{gathered}
\end{equation}
A generic linear combination is not a cubic in the fiber coordinates
$x$, $y$ because of the $w^4 x^2 y^2$ term. This comes as no surprise,
since an anticanonical hypersurface in the $\CP^1\times\CP^1$ fiber is
a biquadric in $x$ and $y$. Of course a smooth\footnote{This is not
  true for singular biquadrics in $\CP^1\times\CP^1$, for example the
  ``large complex structure limit'' $x_0 x_1 y_0 y_1$ has four
  irreducible components, whereas a cubic in $\CP^2$ can have at most
  three.} biquadric in $\CP^1\times\CP^1$ is isomorphic to some cubic
in $\CP^2$, so there \emph{is} a way to bring it into Weierstrass
form. This works as follows~\cite{pre05771645}. Given a biquadric
\begin{equation}
  \label{eq:biquadric}
  q(x,y) = 
  \alpha_{22} x^2 y^2 + \alpha_{21} x^2 y + \alpha_{20} x^2 +
  \alpha_{12} x y^2 + \alpha_{11} x y + x \alpha_{10} + 
  y^2 \alpha_{02} + y \alpha_{01} + \alpha_{00}
  ,
 \end{equation}
first compute the usual quadratic discriminant with respect to $y$, 
\begin{equation}
  \label{eq:biquadricdisc}
  \beta_4 x^4 + \beta_3 x^3 + \beta_2 x^2 + \beta_1 x + \beta_0 =
  \left(
    \sum \alpha_{i1} x^i
  \right)^2
  - 4
  \left(
    \sum \alpha_{i2} x^i
  \right)
  \left(
    \sum \alpha_{i0} x^i
  \right)
  .
\end{equation}
The coefficients $a$, $b$ of the Weierstrass form $y^2=x^3+ax+b$ are
then given by the quadratic and cubic projective $GL(2,\C)$-invariants
of the resulting plane quartic,
\begin{equation}
  \label{eq:WeierstrassBiquadric}
  \begin{split}
    a =& - \tfrac{1}{4} \big(
    \beta_0 \beta_4 + 3 \beta_2^2 - 4 \beta_1 \beta_3
    \big) \\
    b =& - \tfrac{1}{4} \big(
    \beta_0 \beta_3^2 +\beta_1^2 \beta_4 -\beta_0 \beta_2 \beta_4 
    -2 \beta_1 \beta_2 \beta_3 +\beta_2^3
    \big)
    .
  \end{split}
\end{equation}

\subsection{Gauge Group}
\label{sec:gauge}

It is now an easy exercise to bring any chosen Calabi-Yau hypersurface
into Weierstrass form with $35$ free parameters. However, due to the
number of coefficients the result will be unwieldy. For illustration,
we will therefore pick the following ``random'' coefficients for the
monomials in eq.~\eqref{eq:monomials}
\begin{equation}
  \begin{gathered}
    (1,2,-2,1,2,2,1,-2,0,0,2,1,-1,2,2,2,1,-1,
    \hspace{2cm} \\ \hspace{2cm}
    0,0,-1,1,1,0,2,2,2,2,-1,-2,2,2,-1,-1,1)
  \end{gathered}
\end{equation}
\begin{table}
  \centering
  \renewcommand{\arraystretch}{1.5}
  \begin{tabular}{c|cccccccccc}
    Fiber &
    $I_0$ &  
    $I_n$ &
    $II$ &
    $III$ & 
    $IV$ & 
    $I_0^*$ &
    $I_{n}^*$ & 
    $IV^*$ &
    $III^*$ &
    $II^*$
    \\
    \hline
    $\ord(a)$ &
    $\geq 0$  & 
    $0$       & 
    $\geq 0$  & 
    $1$       & 
    $\geq 2$  & 
    $\geq 2$  & 
    $2$       & 
    $\geq 3$  & 
    $3$       & 
    $\geq 4$    
    \\
    $\ord(b)$ &
    $\geq 0$  & 
    $0$       & 
    $1$       & 
    $\geq 2$  & 
    $2$       & 
    $\geq 3$  & 
    $3$       & 
    $4$       & 
    $\geq 5$  & 
    $5$         
    \\
    $\ord(\Delta)$ &
    $0$ & 
    $n$      & 
    $2$      & 
    $3$      & 
    $4$      & 
    $6$      & 
    $n+6$    & 
    $8$      & 
    $9$      & 
    $10$       
  \end{tabular}
  \caption{Tate's algorithm~\cite{MR0393039} for the Kodaira fiber of
    a Weierstrass
    equation.}
  \label{tab:Tate}
\end{table}
We have verified that these are sufficiently random in the sense that
any other generic choice will lead to the same orders of vanishing and
factorizations in the following. The coefficients and discriminant of
the Weierstrass form are, then, 
\begin{equation}
  \begin{split}
    a &=
    -\tfrac{1}{48} \big( 16 u^{12} - 32 u^{10} v^2 + 32 u^9 v^3 
    + \cdots - 71 w^{12} \big)
    \\
    b &=
    \tfrac{1}{864} \big( 64 u^{18} - 192 u^{16} v^2 + 192 u^{15} v^3  
    + \cdots + 269 w^{18} \big)
    \\
    \Delta &= 
    -\tfrac{1}{16} \; w^{10} \;
    \big(1024 u^{26} + 2048 u^{25} v - 11008 u^{24} v^2 
    + \cdots - 249 w^{26} \big)
  \end{split}
\end{equation}
where the factors containing the ellipses are irreducible (and contain
a great number of terms). Using \autoref{tab:Tate}, we can immediately
read off that the discriminant divisor splits into an $I_{10}$
component along the toric divisor ${w=0}\subset \CP^2$ and an $I_1$
component on a degree-$26$ curve.

The low-energy gauge group depends on the Kodaira type along each
discriminant component as well as the monodromy\footnote{The
  discriminant component ${w=0}=\CP^1\subset\CP^2$ is simply
  connected, $\pi_1(\CP^1)=1$. Nevertheless there can (and generally
  will) be a monodromy, because one has to excise the points of
  intersection with the $I_1$ discriminant component.} of the Kodaira
fiber. Whether or not there is a monodromy can also be read off from
the Weierstrass model~\cite{MR0393039, Bershadsky:1996nh, Katz:2011qp,
  Grassi:2011hq}. For the $I_m$ case, $m\geq 3$, this depends on
whether $\tfrac{b}{a}$ restricted to the discriminant is a square or
not. In the case at hand one obtains
\begin{equation}
  \left.
    \frac{b}{a}
  \right|_{w=0} = 
  -\tfrac{1}{18}  (u + v)^2  \big(2 u^2 - 2 u v + v^2\big)^2
  .
\end{equation}
Hence we are in the ``split'' case, and the gauge group is
$SU(10)$. As we will see in the next subsection, the fact that there
is no monodromy can be nicely be seen from the toric geometry of the
resolved Calabi-Yau threefold.

\subsection{Resolution of Singularities}
\label{sec:resolv}

So far, we only discussed the Weierstrass model without going into the
details of the resolution of singularities. Really, this is the
essential novelty of the approach taken in this paper: By starting
from the maximal resolutions of Gorenstein Fano toric varieties, we
have complete control over the desingularization of the Weierstrass
model. In particular, the details of the resolution of singularities
are visible and the Hodge numbers can be readily computed.
\begin{figure}
  \centering
  \includegraphics{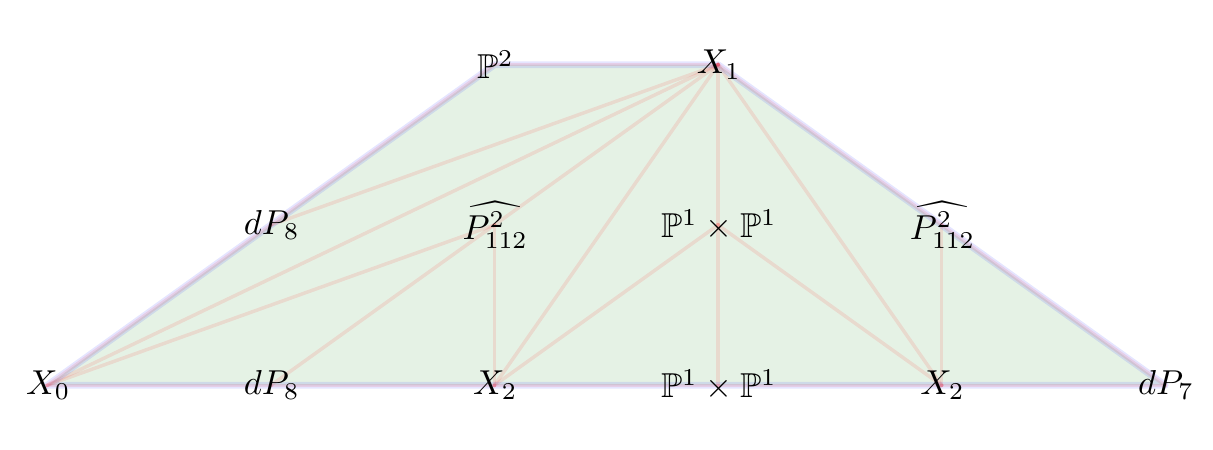}
  \\
  \includegraphics{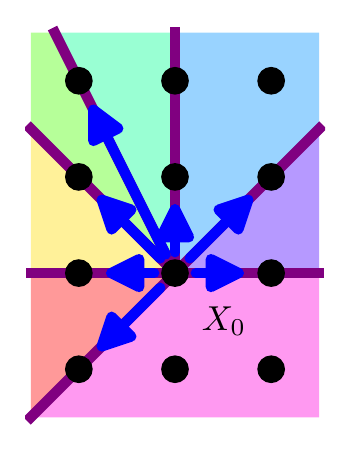}
  \includegraphics{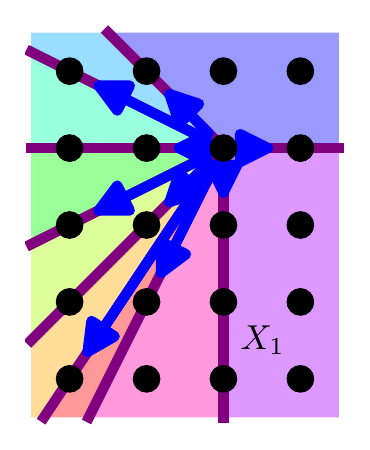}
  \includegraphics{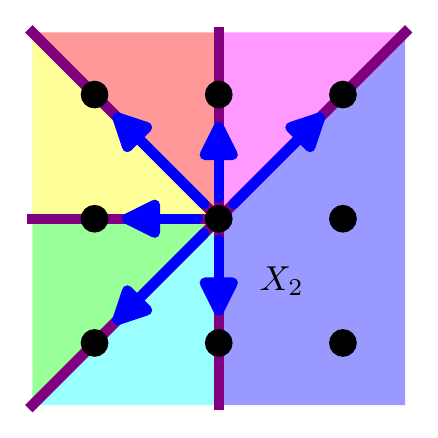}
  \caption{The irreducible fiber components over $w=0$. The top row
    shows the different toric surfaces that form the irreducible
    components. The bottom row consists of the fans defining the
    surfaces $X_0$, $X_1$, and $X_2$ that do not have standard names.}
  \label{fig:Ffiber}
\end{figure}

To crepantly desingularize the toric variety, we need to subdivide the
fan into smooth (that is, simplicial and unimodular) cones using the
rays through all of the integral points of the polytope. In
particular, one has to utilize the remaining $8$ integral points in
the triangulation of $P$. Any such smooth triangulation has 56
generating cones. To be completely explicit, we will be using a
particular triangulation that is uniquely determined by admitting a
toric fibration together with the induced triangulation of the
two-face $F\subset P$ shown in \autoref{fig:Ftriang}. Using this fan,
there are $12$ primitive~\cite{2000math.....10082H} preimage cones
over $\cone{w}$, namely the $12$ integral points of the two-face
$F$. Therefore, the toric fiber over $w=0$ in the total space consists
of $12$ irreducible components. Each irreducible component is a toric
surface, and they are joined along common $\CP^1$ as the corresponding
points of the induced triangulation of $F$. The details of all toric
fiber components are shown in \autoref{fig:Ffiber}.

The Calabi-Yau hypersurface can, but does not have to, intersect the
toric fiber components. To determine the Kodaira type of the
degenerate elliptic fiber, one needs to restrict the anticanonical
divisor on the ambient toric variety to each of the irreducible
components of the toric fiber. One finds that the restriction is
trivial for the two toric fiber components in the interior of the
two-face $F$, and nontrivial for the $10$ toric fiber components
corresponding to the points on the boundary of $F$. This is how the
Kodaira fiber $I_{10}$ arises over the $w=0$ component of the
discriminant in the elliptic fibration:
\begin{itemize}
\item The Calabi-Yau hypersurface intersects each of the $10$
  two-dimensional toric fibers on the boundary of $F$ in a $\CP^1$.
\item There is a one-dimensional fiber component (where two
  $2$-dimensional components intersect) for each of the lines of the
  triangulation of $F$. The Calabi-Yau hypersurface intersects each of
  the $10$ blue lines in \autoref{fig:Ffiber} in a point, and does not
  intersect any of the $13$ red lines.
\end{itemize}
\begin{figure}
  \centering
  \includegraphics{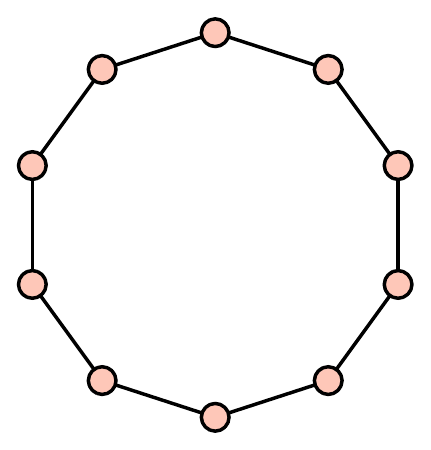}
  \caption{The fiber-divisor-graph $F(\langle w\rangle, -K)$ for the
    fiber over the discriminant component $w=0$. It is the
    $\tilde{A}_9$ extended Dynkin diagram corresponding to an $I_{10}$
    Kodaira fiber.}
  \label{fig:fdgI10}
\end{figure}
Hence, in this example the graph of the $I_{10}$ Kodaira fiber (that
is, the $\tilde{A}_9$ extended Dynkin diagram) is visible as the graph
of integral points on $\partial F$ and their connecting
edges~\cite{Candelas:1996ht, Candelas:1997pv, Candelas:1997eh,
  Candelas:1997pq, Candelas:1997jz, Braun:2000hh}. As we will see in
the next section, this is not always true. However, much of the
information about the Kodaira fiber can be derived from the pull-back
of the anticanonical divisor to the toric fibers. A nice graphical
representation of this data is what we will call the \emph{fiber
  divisor graph} in the following:
\begin{definition}[Fiber-Divisor-Graph]
  Let $\pi:X\to B$ be a toric fibration with $2$-dimensional fibers
  and $\dim(X)=4$. For a fixed toric fiber $\pi^{-1}(p) = \cup F_i$
  and nef divisor $D\subset X$, let $F(p,D)$ be the graph with
  \begin{itemize}
  \item one node for each fiber irreducible component $F_i$ such that
    $[D|_{F_i}]\not=0$, and
  \item $\Z_{\geq 0} \owns D\cap F_i \cap F_j$ edges joining $F_i$ and $F_j$.
  \end{itemize} 
  The fiber-divisor-graph only depends on the torus orbit of the base
  point $p\in B$ (that is, a cone of the fan of $B$) and the divisor
  class $[D]$.
\end{definition}
For example, the fiber-divisor-graph $F(\langle w\rangle, -K)$ for the
example discussed in this section is shown in \autoref{fig:fdgI10}.


\section{Non-Split Fibrations}
\label{sec:split}

\subsection{Weierstrass Model}

We now turn to a more complicated example that will explain how to
deal with various issues in classifying the gauge groups of toric
elliptic fibrations. Apart from the origin, the polytope contains the
integral points in the following table:
\begin{equation}
  \begin{array}{|cccccc|ccccccccc|cc|}
    \hline
    s_0&s_1&s_2&s_3&s_4&s_5&s_6&s_7
    &\multicolumn{9}{c|}{}
    \\ \hline
    t_0&t_1&t_2&t_3&t_4&
    t_5&t_6&t_7&t_8&t_9&
    t_{10}&t_{11}&t_{12}&t_{13}&t_{14}&t_{15}&t_{16}
    \\ \hline
    -2 & -2 & 0 & 0 & 1 & 0 & 0 & 0 & -2 & -1 & -1 & -1 
    & 0 & 0 & 0 & -1 & 0 \\
    -2 & -2 & 1 & 1 & 0 & 0 & 0 & 0 & -2 & -1 & -1 & -1 
    & 1 & 1 & 1 & -1 & 0 \\
    -5 & -5 & 0 & 0 & 0 & 1 & -1 & -1 & -5 & -3 & -3 & -2 
    & 0 & 0 & 0 & -3 & -1 \\
    -5 & -3 & 0 & 4 & 0 & 0 & -1 & 1 & -4 & -3 & -1 & -2 
    & 1 & 2 & 3 & -2 & 0 \\
    \hline
    \multicolumn{6}{|c|}{\text{vertices}} & 
    \multicolumn{9}{c|}{\text{not interior to facets}} & 
    \multicolumn{2}{c|}{\text{rest}}
    \\
    \hline
  \end{array}
\end{equation}
The first 6 points are the vertices, the middle 9 points are integral
points that are not interior to facets, and the last 2 points are
interior to facets. The eight $s_i$ are the homogeneous coordinates
necessary to write the Weierstrass model; the seventeen $t_j$ are the
homogeneous coordinates necessary to completely desingularize the
elliptically fibered Calabi-Yau hypersurface and will only play a role
in the next subsection.

The most coarse toric variety would use only the
vertices as rays of the fan, but this alone is not sufficient for a
toric fibration. In particular, we will use the toric morphism defined
by the projection onto the first two coordinates, that is,
\begin{equation}
  \varphi(\vec{n}) = 
  \begin{pmatrix}
    1 & 0 & 0 & 0 \\
    0 & 1 & 0 & 0 \\
  \end{pmatrix}
  \vec{n}
\end{equation}
A minimal subdivision of the face fan for which $\varphi$ does define
a toric fibration is generated by the following $16$ four-dimensional
cones:
\begin{equation}
  \begin{gathered}
    \smash{\Big\{}
    \langle s_0, s_1, s_2, s_3, s_6, s_7\rangle, 
    \langle s_0, s_1, s_4, s_6, s_7\rangle, 
    \langle s_0, s_1, s_2, s_3, s_5\rangle, 
    \langle s_0, s_1, s_4, s_5\rangle, 
    \hfill \\
    \langle s_0, s_2, s_5, s_6\rangle, 
    \langle s_0, s_4, s_5, s_6\rangle, 
    \langle s_2, s_3, s_4, s_6, s_7\rangle,
    \langle s_1, s_3, s_5, s_7\rangle, 
    \\ \hfill
    \langle s_1, s_4, s_5, s_7\rangle,
    \langle s_2, s_4, s_5, s_6\rangle,
    \langle s_2, s_3, s_4, s_5\rangle, 
    \langle s_3, s_4, s_5, s_7\rangle
    \smash{\Big\}}
  \end{gathered}
\end{equation}
In order to write the Weierstrass form on the maximal torus, we need
to pick coordinates. A slight complication is that there is no ray
whose generator maps onto the generator $\vec{w}$ of the base $\CP^2$
fan, see \autoref{fig:P2}. We only have $\vec{s}_0$ and $\vec{s}_1$ at
our disposal, and both map to $2 \vec{w}$. Hence a choice of
coordinates that map to the base homogeneous coordinates necessarily
involves square roots, for example
\begin{equation}
  \autoref{eq:toricmorphismroot}
  [s_0:\cdots:s_7] = 
  [\sqrt{w}:1:v:1:u:1:1:1]
\end{equation}
Written in terms of $u,v,w$, the hypersurface equation will contain
fractional powers of $w$, but the Weierstrass form will be
polynomial. The dual polytope contains $44$ integral points, so there
are $44$ monomials in the Calabi-Yau hypersurface equation. The
generic fiber is the weighted projective space $\CP^2[1,1,2]$, for
which we explain in \autoref{sec:Weierstrass} how to compute the
Weierstrass form. The result is that
\begin{equation}
  a = P_{10}(u,v,w)
  ,\qquad
  b = P_{15}(u,v,w)
  ,\qquad
  \Delta = v^8 w^2 P_{20}(u,v,w)
  ,
\end{equation}
where $P_d$ is an irreducible polynomial of degree $d$. Hence, the
elliptic fiber over $u=0$ is a smooth elliptic curve, the fiber over
$v=0$ is an $I_8$ Kodaira fiber. Depending on the monodromy of this
$I_8$ fiber, the gauge group can be $SU(8)$ or $Sp(4)$. In this case,
one finds that the monodromy cover $\tfrac{b}{a}\big|_{v=0}$ is not a
square\footnote{Note that one only needs to compute $\gcd\big(p,
  \tfrac{\partial p}{\partial x_1}\big) = 1$, this then guarantees
  that the multivariate polynomial $p(x_1,x_2,\dots)$ is not a
  power. In particular, one does not have to find the splitting
  field.}, hence it is of non-split type.

For the fiber over $w=0$, one needs to be more careful. Clearly the
discriminant vanishes to second order in $w$, but the good local
ambient space coordinate is $\pm\sqrt{w}$. Hence the corresponding
Kodaira fiber is not $I_2$ but $I_2^*$. This can also be derived by
direct computation if one resolves the fan further, for example the
desingularization to be discussed in the following subsection adds new
rays such that some ray generator (for example, $\vec{t}_9$) now maps
onto $\vec{w}$. So by using $t_9$ instead of $s_0$ as local coordinate
on the ambient toric variety, the toric morphism can be written in
terms of polynomials and one obtains the expected Weierstrass form
\begin{equation}
  a = w^2 P_{10}(u,v,w)
  ,\qquad
  b = w^3 P_{15}(u,v,w)
  ,\qquad
  \Delta = v^8 w^8 P_{20}(u,v,w)
  .
\end{equation}
Finally, the monodromy cover $\tfrac{\Delta}{w^8}\big(\tfrac{a
  w}{b}\big)^2|_{w=0} = \psi^2$ factors into the square of a
polynomial, so the $I_2^*$ component is of split type.

To summarize, the three toric divisors on the base $\CP^3$ support the
following gauge groups:
\begin{descriptionlist}
\item{$u=0:$} Elliptic fiber is smooth (Kodaira fiber $I_0$), no
  gauge group.
\item{$v=0:$} Kodaira fiber of type $I_8$, non-split, gauge group
  $Sp(4)$.
\item{$w=0:$} Kodaira fiber of type $I_2^*$, split, gauge group
  $SO(12)$.
\end{descriptionlist}

\subsection{Resolution of Singularities}

The Weierstrass model is just a singular model for the smooth elliptic
fibration in the sense of the minimal model program. We now
desingularize the ambient toric variety, which resolves the Calabi-Yau
hypersurface into a smooth threefold with Hodge numbers
$(h^{11},h^{21})=(19,35)$. This amounts to subdividing the fan until
all cones of dimension $\leq 3$ are smooth. In the following, we will
be using the resolution of the fan generated by the $56$ cones
\begin{equation}
  \label{eq:resolvedfan}
  \begin{gathered}
    \scriptstyle
    \langle t_0, t_2, t_5, t_9\rangle, 
    \langle t_0, t_2, t_5, t_{12}\rangle,
    \langle t_0, t_2, t_9, t_{15}\rangle,
    \langle t_0, t_2, t_{12}, t_{15}\rangle,
    \langle t_0, t_3, t_5, t_{11}\rangle,
    \langle t_0, t_3, t_5, t_{14}\rangle,
    \langle t_0, t_3, t_8, t_{11}\rangle,
    \langle t_0, t_3, t_8, t_{15}\rangle,
    \\
    \scriptstyle
    \langle t_0, t_3, t_{14}, t_{15}\rangle,
    \langle t_0, t_4, t_5, t_9\rangle,
    \langle t_0, t_4, t_5, t_{11}\rangle,
    \langle t_0, t_4, t_8, t_{11}\rangle, 
    \langle t_0, t_4, t_8, t_{15}\rangle,
    \langle t_0, t_4, t_9, t_{15}\rangle,
    \langle t_0, t_5, t_{12}, t_{13}\rangle,
    \langle t_0, t_5, t_{13}, t_{14}\rangle,
    \\
    \scriptstyle
    \langle t_0, t_{12}, t_{13}, t_{15}\rangle,
    \langle t_0, t_{13}, t_{14}, t_{15}\rangle,
    \langle t_1, t_3, t_5, t_{10}\rangle, 
    \langle t_1, t_3, t_5, t_{11}\rangle,
    \langle t_1, t_3, t_8, t_{10}\rangle,
    \langle t_1, t_3, t_8, t_{11}\rangle,
    \langle t_1, t_4, t_5, t_{10}\rangle,
    \langle t_1, t_4, t_5, t_{11}\rangle,
    \\
    \scriptstyle
    \langle t_1, t_4, t_8, t_{10}\rangle,
    \langle t_1, t_4, t_8, t_{11}\rangle,
    \langle t_2, t_4, t_5, t_6\rangle,
    \langle t_2, t_4, t_5, t_{12}\rangle,
    \langle t_2, t_4, t_6, t_{16}\rangle,
    \langle t_2, t_4, t_7, t_{12}\rangle, 
    \langle t_2, t_4, t_7, t_{16}\rangle,
    \langle t_2, t_5, t_6, t_9\rangle,
    \\
    \scriptstyle
    \langle t_2, t_6, t_9, t_{16}\rangle,
    \langle t_2, t_7, t_9, t_{15}\rangle,
    \langle t_2, t_7, t_9, t_{16}\rangle,
    \langle t_2, t_7, t_{12}, t_{15}\rangle,
    \langle t_3, t_4, t_5, t_7\rangle, 
    \langle t_3, t_4, t_5, t_{14}\rangle,
    \langle t_3, t_4, t_7, t_{14}\rangle,
    \langle t_3, t_5, t_7, t_{10}\rangle,
    \\
    \scriptstyle
    \langle t_3, t_7, t_8, t_{10}\rangle,
    \langle t_3, t_7, t_8, t_{15}\rangle,
    \langle t_3, t_7, t_{14}, t_{15}\rangle, 
    \langle t_4, t_5, t_6, t_9\rangle, 
    \langle t_4, t_5, t_7, t_{10}\rangle,
    \langle t_4, t_5, t_{12}, t_{13}\rangle,
    \langle t_4, t_5, t_{13}, t_{14}\rangle,
    \langle t_4, t_6, t_9, t_{16}\rangle, 
    \\
    \scriptstyle
    \langle t_4, t_7, t_8, t_{10}\rangle, 
    \langle t_4, t_7, t_8, t_{15}\rangle, 
    \langle t_4, t_7, t_9, t_{15}\rangle, 
    \langle t_4, t_7, t_9, t_{16}\rangle,
    \langle t_4, t_7, t_{12}, t_{13}\rangle,
    \langle t_4, t_7, t_{13}, t_{14}\rangle, 
    \langle t_7, t_{12}, t_{13}, t_{15}\rangle,
    \langle t_7, t_{13}, t_{14}, t_{15}\rangle.
  \end{gathered}
\end{equation}
The corresponding toric variety still has point-like orbifold
singularities, but they will be missed by a generic Calabi-Yau
hypersurface.

It is a subtle point that we need to add the rays through the points
$\vec{t}_{15}$ and $\vec{t}_{16}$ that are in the interior of a facet
of the polytope to resolve the toric variety to be smooth except for
point singularities and, \emph{at the same time}, be fibered over
$\CP^2$ by $\varphi$. If we would not require the fibration structure,
we could just merge the generating cones containing $\vec{t}_{15}$,
$\vec{t}_{16}$, that is, replace
\begin{equation}
  \begin{split}
    \left\{
      \begin{smallmatrix}
        \langle t_0, t_3, t_8, t_{15}\rangle,
        \langle t_3, t_7, t_8, t_{15}\rangle,
        \langle t_4, t_7, t_8, t_{15}\rangle,
        \langle t_0, t_4, t_8, t_{15}\rangle,
        \\
        \langle t_4, t_7, t_9, t_{15}\rangle,
        \langle t_0, t_4, t_9, t_{15}\rangle,
        \langle t_2, t_7, t_9, t_{15}\rangle,
        \langle t_0, t_2, t_9, t_{15}\rangle,
        \\
        \langle t_0, t_2, t_{12}, t_{15}\rangle,
        \langle t_2, t_7, t_{12}, t_{15}\rangle,
        \langle t_0, t_{12}, t_{13}, t_{15}\rangle,
        \langle t_7, t_{12}, t_{13}, t_{15}\rangle,
        \\
        \langle t_0, t_3, t_{14}, t_{15}\rangle,
        \langle t_3, t_7, t_{14}, t_{15}\rangle,
        \langle t_0, t_{13}, t_{14}, t_{15}\rangle,
        \langle t_7, t_{13}, t_{14}, t_{15}\rangle
      \end{smallmatrix}\right\}
    \longrightarrow&~
    \langle t_0, t_2, t_3, t_4, t_7, t_8, t_9, t_{12}, t_{13},
    t_{14}\rangle
    ,
    \\
    \left\{
      \begin{smallmatrix}
        \langle t_2, t_4, t_6, t_{16}\rangle,
        \langle t_2, t_4, t_7, t_{16}\rangle,
        \langle t_4, t_7, t_9, t_{16}\rangle,
        \\
        \langle t_4, t_6, t_9, t_{16}\rangle,
        \langle t_2, t_7, t_9, t_{16}\rangle,
        \langle t_2, t_6, t_9, t_{16}
      \end{smallmatrix}\right\}
    \longrightarrow&~
    \langle t_2, t_4, t_6, t_7, t_9\rangle
    .
  \end{split}
\end{equation}
The resulting toric variety would still only have point singularities,
but would no longer be torically fibered.
\begin{figure}[htb]
  \centering
  \vspace{-1cm}
  \includegraphics{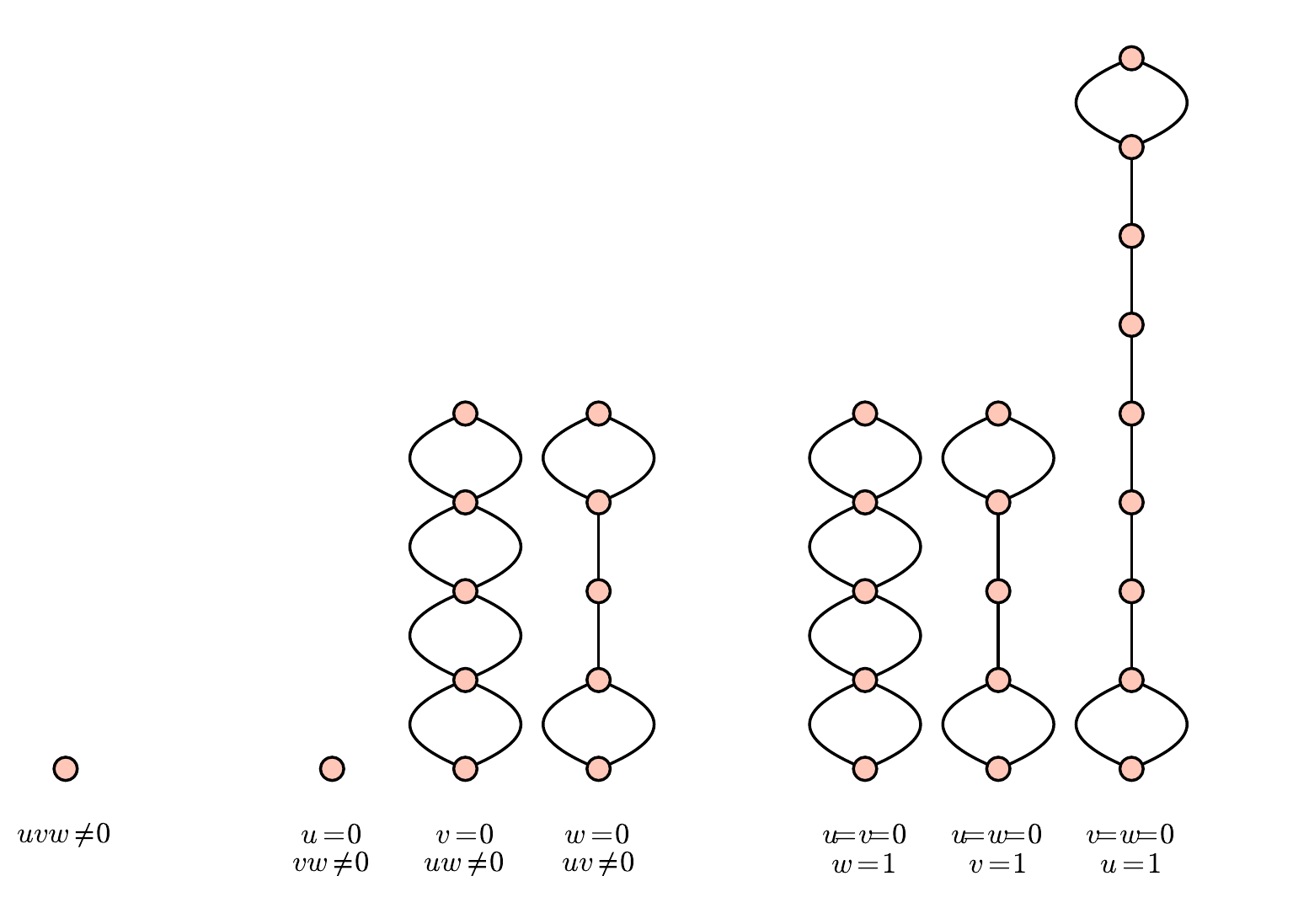}
  \caption{The fiber-divisor-graph $F(O(\sigma), -K)$ of the $I_8
    I_2^*$-fibration over the $7$ torus-orbits $O(\sigma)$, $\sigma
    \in \Sigma$, in the base $\CP^2=\CP_\Sigma$.}
  \label{fig:fdgNonSplit}
\end{figure}

\begin{figure}
  \centering
  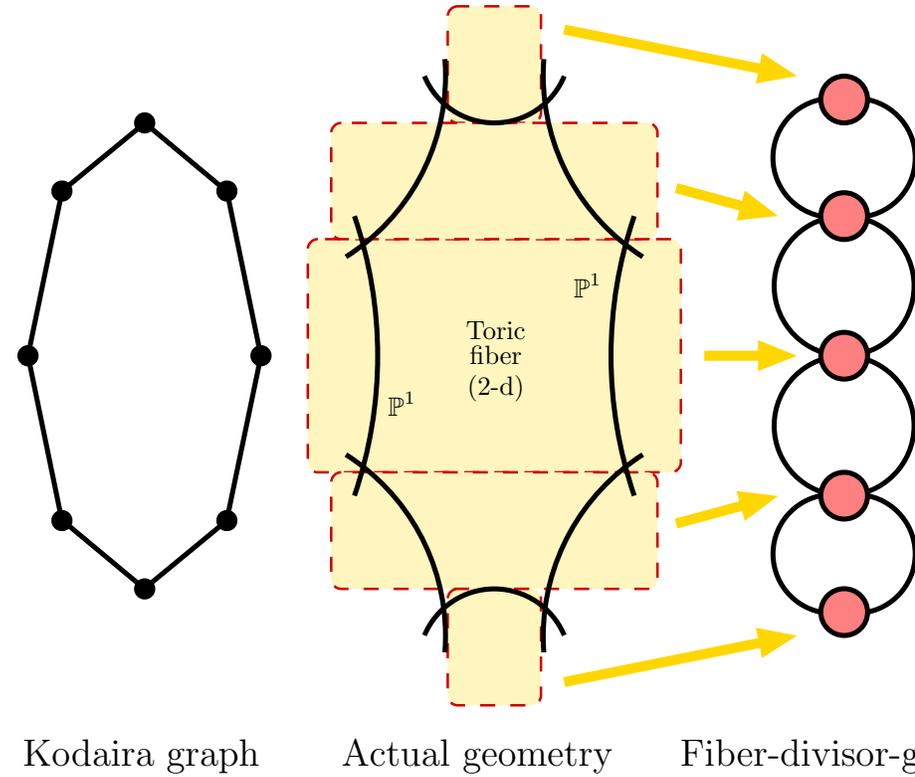
  \caption{Relation between the different visualizations of the
    degenerate fiber over the $\{v=0\}$ discriminant component. The
    actual geometry of the $I_8$ Kodaira fiber consists of $8$ $\CP^1$
    intersecting in a ring. They are contained in $5$ different
    irreducible components of the 2-dimensional toric fiber. The
    fiber-divisor-graph contracts each component of the toric fiber to
    a node, joined by an edge for each intersection of the contained
    $\CP^1$. The Kodaira graph is the graph dual graph to the $8$
    $\CP^1$, ignoring the embedding in the toric fiber.}
  \label{fig:fibergraphs}
\end{figure}

Using the resolved fan eq.~\eqref{eq:resolvedfan}, it is now a
straightforward exercise to compute the restriction of the
anticanonical divisor to each toric fiber. The fiber-divisor-graph
introduced in \autoref{sec:resolv} is a useful way of visualizing the
result, and can be seen in \autoref{fig:fdgNonSplit}. One immediately
notices that the graphs of the fibers over $v=0$ and $w=0$ do not look
like the graphs of the expected $I_8$ and $I_2^*$ Kodaira fibers. In
fact, the fiber-divisor-graph over $v=0$ cannot be the $\tilde{A}_{7}$
extended Dynkin diagram. This is because the irreducible components of
the fibers of a toric morphism undergo no monodromy. Hence, if the
$I_{8}$ Kodaira fiber were realized by eight $\CP^1$s in eight
different irreducible components of the toric fiber, the $\CP^1$
components would be locked in place and could not undergo any
monodromy either. Hence, the discriminant component would necessarily
be of split type! In other words, a non-split discriminant component
requires that some irreducible component of the toric fiber contains
multiple disjoint $\CP^1$s, which can then be exchanged by monodromies
of the hypersurface equation. See \autoref{fig:fibergraphs} for a
visualization of how the fiber geometry determines the
fiber-divisor-graph.

To summarize, we now see how the geometry of the degenerate fiber is
encoded in the fiber-divisor-graph. In the split case, the graph can
be equal to the associated extended Dynkin diagram, but in general (in
particular, in the non-split case) it arises from identifying nodes of
the extended Dynkin diagram that correspond to $\CP^1$ embedded in the
same irreducible toric fiber component. The example discussed in this
section is a Miranda fibration~\cite{MR690264} where a $I_8$ and an
$I_2^*$ component of the discriminant intersect transversely. The
degenerate fiber over the intersection point $[u:v:w]=[0:1:0]$ of the
two discriminant components is an $I_6^*$ Kodaira fiber, see
\autoref{fig:fdgNonSplit}, as expected from a Miranda fibration.


\section{Non-Flat Fibrations}
\label{sec:nonflat}

Finally, let quickly go through an example of a non-flat fibration. As
we already mentioned, these form the bulk of all fibrations over
$\CP^2$, though they should more properly be studied as fibrations
over a blown-up base. Apart from the origin, the fibered reflexive
polytope contains the $19$ integral points 
\begin{equation}
  \setlength\arraycolsep{1.35mm} 
   \begin{array}{|ccccccc|cccccccccccc|}
     \hline
     z_0&z_1&z_2&z_3&z_4&
     z_5&z_6&z_7&z_8&z_9&
     z_{10}&z_{11}&z_{12}&z_{13}&z_{14}&
     z_{15}&z_{16}&z_{17}&z_{18}
     \\ \hline
     -3 & 0 & 0 & 0 & 0 & 0 & 1 & -2 & -2 & -1 & -1 & -1 
     & -1 & -1 & 0 & 0 & 0 & 0 & 0 \\
     -3 & 3 & 0 & 1 & 0 & 0 & 0 & -1 & -2 & 1 & 0 & 0 
     & -1 & -1 & 2 & 2 & 1 & 1 & 1 \\
     3 & -6 & -1 & 0 & 0 & 1 & 0 & 0 & 2 & -3 & -1 & 0 
     & 1 & 1 & -4 & -3 & -2 & -2 & -1 \\
     1 & -2 & -1 & 0 & 1 & 0 & 0 & 0 & 1 & -1 & 0 & 0 
     & 0 & 1 & -1 & -1 & -1 & 0 & 0 \\
     \hline
     \multicolumn{7}{|c|}{\text{vertices}} & 
     \multicolumn{12}{c|}{\text{remaining integral points}}
     \\
     \hline
  \end{array}
\end{equation}
and is fibered by the sub-polytope $\conv\{\vec{z}_2, \vec{z}_4,
\vec{z}_5\}$, or, equivalently, by the lattice projection
\begin{equation}
  \varphi(\vec{n}) = 
  \begin{pmatrix}
    1 & 0 & 0 & 0 \\
    0 & 1 & 0 & 0
  \end{pmatrix}
  \vec{n}
\end{equation}
The naive geometric image of the rays through the $19$ integral points
contains the rays generated by $(-2,-1)$, $(-1,0)$, and $(-1,1)$ in
addition to the rays of the fan of $\CP^2$, \autoref{fig:P2}, showing
that this cannot be a \emph{flat} fibration over $\CP^2$.

Nevertheless, we can easily construct a (non-flat) fibration of a
toric variety over $\CP^2$. We take the total space fan to be
generated by the $66$ cones
\begin{equation}
  \begin{gathered}
    \scriptstyle
    \langle z_0, z_2, z_6, z_8\rangle,
    \langle z_0, z_2, z_6, z_{12}\rangle,
    \langle z_0, z_2, z_7, z_8\rangle,
    \langle z_0, z_2, z_7, z_{12}\rangle,
    \langle z_0, z_3, z_5, z_8\rangle,
    \langle z_0, z_3, z_5, z_{11}\rangle,
    \langle z_0, z_3, z_8, z_{11}\rangle,
    \\
    \scriptstyle
    \langle z_0, z_5, z_6, z_8\rangle,
    \langle z_0, z_5, z_6, z_{12}\rangle,
    \langle z_0, z_5, z_7, z_{11}\rangle,
    \langle z_0, z_5, z_7, z_{12}\rangle,
    \langle z_0, z_7, z_8, z_{11}\rangle,
    \langle z_1, z_2, z_6, z_{14}\rangle,
    \langle z_1, z_2, z_6, z_{16}\rangle,
    \\
    \scriptstyle
    \langle z_1, z_2, z_9, z_{14}\rangle,
    \langle z_1, z_2, z_9, z_{16}\rangle,
    \langle z_1, z_5, z_6, z_{15}\rangle,
    \langle z_1, z_5, z_6, z_{16}\rangle,
    \langle z_1, z_5, z_9, z_{15}\rangle,
    \langle z_1, z_5, z_9, z_{16}\rangle,
    \langle z_1, z_6, z_{14}, z_{15}\rangle,
    \\
    \scriptstyle
    \langle z_1, z_9, z_{14}, z_{15}\rangle,
    \langle z_2, z_4, z_6, z_{13}\rangle,
    \langle z_2, z_4, z_6, z_{17}\rangle,
    \langle z_2, z_4, z_7, z_{10}\rangle,
    \langle z_2, z_4, z_7, z_{13}\rangle,
    \langle z_2, z_4, z_9, z_{10}\rangle,
    \langle z_2, z_4, z_9, z_{17}\rangle,
    \\
    \scriptstyle
    \langle z_2, z_5, z_6, z_{12}\rangle,
    \langle z_2, z_5, z_6, z_{16}\rangle,
    \langle z_2, z_5, z_7, z_9\rangle,
    \langle z_2, z_5, z_7, z_{12}\rangle,
    \langle z_2, z_5, z_9, z_{16}\rangle,
    \langle z_2, z_6, z_8, z_{13}\rangle,
    \langle z_2, z_6, z_{14}, z_{17}\rangle,
    \\
    \scriptstyle
    \langle z_2, z_7, z_8, z_{13}\rangle,
    \langle z_2, z_7, z_9, z_{10}\rangle,
    \langle z_2, z_9, z_{14}, z_{17}\rangle,
    \langle z_3, z_4, z_5, z_6\rangle,
    \langle z_3, z_4, z_5, z_{13}\rangle,
    \langle z_3, z_4, z_6, z_{18}\rangle,
    \langle z_3, z_4, z_9, z_{10}\rangle,
    \\
    \scriptstyle
    \langle z_3, z_4, z_9, z_{18}\rangle,
    \langle z_3, z_4, z_{10}, z_{11}\rangle,
    \langle z_3, z_4, z_{11}, z_{13}\rangle,
    \langle z_3, z_5, z_6, z_{15}\rangle,
    \langle z_3, z_5, z_8, z_{13}\rangle,
    \langle z_3, z_5, z_9, z_{11}\rangle,
    \langle z_3, z_5, z_9, z_{15}\rangle,
    \\
    \scriptstyle
    \langle z_3, z_6, z_{14}, z_{15}\rangle,
    \langle z_3, z_6, z_{14}, z_{18}\rangle,
    \langle z_3, z_8, z_{11}, z_{13}\rangle,
    \langle z_3, z_9, z_{10}, z_{11}\rangle,
    \langle z_3, z_9, z_{14}, z_{15}\rangle,
    \langle z_3, z_9, z_{14}, z_{18}\rangle,
    \\
    \scriptstyle
    \langle z_4, z_5, z_6, z_{13}\rangle,
    \langle z_4, z_6, z_{17}, z_{18}\rangle,
    \langle z_4, z_7, z_{10}, z_{11}\rangle,
    \langle z_4, z_7, z_{11}, z_{13}\rangle,
    \langle z_4, z_9, z_{17}, z_{18}\rangle,
    \langle z_5, z_6, z_8, z_{13}\rangle,
    \\
    \scriptstyle
    \langle z_5, z_7, z_9, z_{11}\rangle,
    \langle z_6, z_{14}, z_{17}, z_{18}\rangle,
    \langle z_7, z_8, z_{11}, z_{13}\rangle,
    \langle z_7, z_9, z_{10}, z_{11}\rangle,
    \langle z_9, z_{14}, z_{17}, z_{18}\rangle.
  \end{gathered}
\end{equation}
The 4-dimensional toric variety is smooth apart from isolated orbifold
singularities, so a generic Calabi-Yau hypersurface will be a smooth
threefold with Hodge numbers $(h^{11},h^{21})=(19,25)$. The fibration
is not flat because some 1-cones (rays) of the ambient toric variety
map to the interior of 2-cones of the base fan. Note, however, that
one cannot simply blow up the base (that is, subdivide the $\CP^2$
fan) and still retain a fibration: There are a number of 2-cones in
the 4-d fan that map onto the three 2-cones of the base, for example
\begin{equation}
  \varphi\big( \langle z_3, z_6 \rangle \big) = \langle u, v \rangle
  ,\qquad
  \varphi\big( \langle z_0, z_3 \rangle \big) = \langle u, w \rangle
  ,\qquad
  \varphi\big( \langle z_0, z_6 \rangle \big) = \langle v, w \rangle
  .  
\end{equation}
Hence, if one wanted to flatten the fibration by blowing up the base,
one would first have to perform flop transitions on the ambient toric
variety corresponding to bistellar flips that eliminate these
offending cones. This can always be done, but will not be the subject
of this section.

We proceed to pick coordinates on the maximal torus
\begin{equation}
  [z_0:\cdots:z_{18}] 
  = 
  [1: 1: z: v: y: x: u: 1: 1: 1: 1: 1: w: 1: 1: 1: 1: 1: 1]
  .
\end{equation}
In this patch the Calabi-Yau hypersurface equation reads
\begin{equation}
  \begin{gathered}
  p(u,v,w,x,y,z) = 
  a_{0} v^4 w x^3 + a_{1} v^3 w x^2 y + a_{2} u v^2 w x^2 z 
  + a_{3} v^3 w x^2 z + a_{4} v^2 w^2 x^2 z 
  \hfill
  \\
  + a_{5} v^2 w x y^2 
  + a_{6} u v w x y z + a_{7} v^2 w x y z + a_{8} v w^2 x y z 
  + a_{9} u^2 w x z^2 + a_{10} u v w x z^2 
  \\
  + a_{11} v^2 w x z^2 
  + a_{12} u w^2 x z^2 + a_{13} v w^2 x z^2 + a_{14} w^3 x z^2 
  + a_{15} u^2 y^3 + a_{16} u v y^3 
  \\
  + a_{17} v^2 y^3 
  + a_{18} u w y^3 
  + a_{19} v w y^3 + a_{20} w^2 y^3 + a_{21} u w y^2 z 
  + a_{22} v w y^2 z + a_{23} w^2 y^2 z 
  \\ 
  \hfill
  + a_{24}u w y z^2 
  + a_{25} v w y z^2 + a_{26} w^2 y z^2 + a_{27} u w z^3 
  + a_{28} v w z^3 + a_{29} w^2 z^3.
  \end{gathered}
\end{equation}
Since the fiber polytope was just the polytope of $\CP^2$, the
equation is a cubic in $[x:y:z]$. Transforming it into Weierstrass
form, one obtains
\begin{equation}
  a = v^3 w^3 P_{6}(u,v,w)
  ,\qquad
  b = v^4 w^4 P_{10}(u,v,w)
  ,\qquad
  \Delta = v^8 w^8 P_{20}(u,v,w)
  ,
\end{equation}
so the discriminant consists of two $IV^*$ components over $v=0$ and
$w=0$ as well as an $I_1$ over $P_{20}=0$. The equations for the
monodromy covers are
\begin{equation}
  \begin{split}
    \psi_v^2 = \tfrac{b}{v^4}\big|_{v=0} =&
    \tfrac{1}{4}
    w^4   
    \big(a_{29} u^2 + a_{24} u w + a_{9} w^2\big)^2   
    \big(a_{28} u^2 + a_{22} u w + a_{6} w^2\big)^2   
    \\ &
    \big(a_{27}^2 u^2 - 4 a_{18} a_{29} u^2 - 4 a_{18} a_{24} u w 
    + 2 a_{15} a_{27} u w + a_{15}^2 w^2 
    - \mathrlap{ 4 a_{9} a_{18} w^2 \big),}
    \\[1ex]
    \psi_w^2 = \tfrac{b}{v^4}\big|_{v=0} =&
    \tfrac{1}{4}   
    v^4 
    \big(a_{15} u^2 + a_{16} u v + a_{17} v^2\big)^2
    \\ &     \scriptstyle
    \big(a_2^2 a_9^2 u^6 - 4 a_0 a_9^3 u^6 + 2 a_2 a_3 a_9^2 u^5 v 
    + 2 a_2^2 a_9 a_{10} u^5 v - 12 a_0 a_9^2 a_{10} u^5 v 
    + a_3^2 a_9^2 u^4 v^2 
    \\[-1ex] & \phantom{\big(} \scriptstyle
    + 4 a_2 a_3 a_9 a_{10} u^4 v^2 
    + a_2^2 a_{10}^2 u^4 v^2 - 12 a_0 a_9 a_{10}^2 u^4 v^2 
    + 2 a_2^2 a_9 a_{11} u^4 v^2 
    - 12 a_0 a_9^2 a_{11} u^4 v^2 
    \\[-1ex] & \phantom{\big(} \scriptstyle
    - 4 a_2^3 a_{27} u^4 v^2 + 18 a_0 a_2 a_9 a_{27} u^4 v^2 
    + 2 a_3^2 a_9 a_{10} u^3 v^3 + 2 a_2 a_3 a_{10}^2 u^3 v^3 
    - 4 a_0 a_{10}^3 u^3 v^3
    \\[-1ex] & \phantom{\big(} \scriptstyle
    + 4 a_2 a_3 a_9 a_{11} u^3 v^3 
    + 2 a_2^2 a_{10} a_{11} u^3 v^3 - 24 a_0 a_9 a_{10} a_{11} u^3 v^3 
    - 12 a_2^2 a_3 a_{27} u^3 v^3 
    \\[-1ex] & \phantom{\big(} \scriptstyle
    + 18 a_0 a_3 a_9 a_{27} u^3 v^3 
    + 18 a_0 a_2 a_{10} a_{27} u^3 v^3 - 4 a_2^3 a_{28} u^3 v^3 
    + 18 a_0 a_2 a_9 a_{28} u^3 v^3
    + a_3^2 a_{10}^2 u^2 v^4 
    \\[-1ex] & \phantom{\big(} \scriptstyle
    + 2 a_3^2 a_9 a_{11} u^2 v^4 + 4 a_2 a_3 a_{10} a_{11} u^2 v^4 
    - 12 a_0 a_{10}^2 a_{11} u^2 v^4 
    + a_2^2 a_{11}^2 u^2 v^4 
    - 12 a_0 a_9 a_{11}^2 u^2 v^4 
    \\[-1ex] & \phantom{\big(} \scriptstyle
    - 12 a_2 a_3^2 a_{27} u^2 v^4 
    + 18 a_0 a_3 a_{10} a_{27} u^2 v^4 
    + 18 a_0 a_2 a_{11} a_{27} u^2 v^4 
    - 27 a_0^2 a_{27}^2 u^2 v^4
    \\[-1ex] & \phantom{\big(} \scriptstyle
    - 12 a_2^2 a_3 a_{28} u^2 v^4 
    + 18 a_0 a_3 a_9 a_{28} u^2 v^4 + 18 a_0 a_2 a_{10} a_{28} u^2 v^4 
    + 2 a_3^2 a_{10} a_{11} u v^5 
    \\[-1ex] & \phantom{\big(} \scriptstyle
    + 2 a_2 a_3 a_{11}^2 u v^5 
    - 12 a_0 a_{10} a_{11}^2 u v^5 - 4 a_3^3 a_{27} u v^5 
    + 18 a_0 a_3 a_{11} a_{27} u v^5 - 12 a_2 a_3^2 a_{28} u v^5 
    \\[-1ex] & \phantom{\big(} \scriptstyle
    + 18 a_0 a_3 a_{10} a_{28} u v^5
    + 18 a_0 a_2 a_{11} a_{28} u v^5 
    - 54 a_0^2 a_{27} a_{28} u v^5 + a_3^2 a_{11}^2 v^6 
    - 4 a_0 a_{11}^3 v^6
    \\[-1ex] & \phantom{\big(} \scriptstyle
    - 4 a_3^3 a_{28} v^6 
    + 18 a_0 a_3 a_{11} a_{28} v^6 - 27 a_0^2 a_{28}^2 v^6\big)
    ,
  \end{split}
\end{equation}
from which we note that both $IV^*$ components are non-split, leading
to a low-energy $F_4$ (instead of $E_6$) gauge theory. Also, the
monodromy cover breaks the exchange symmetry between the two gauge
groups that one might have naively expected.\footnote{This was to be
  expected as the defining polytope has no symmetries.}

\begin{figure}[htb]
  \centering
  \includegraphics{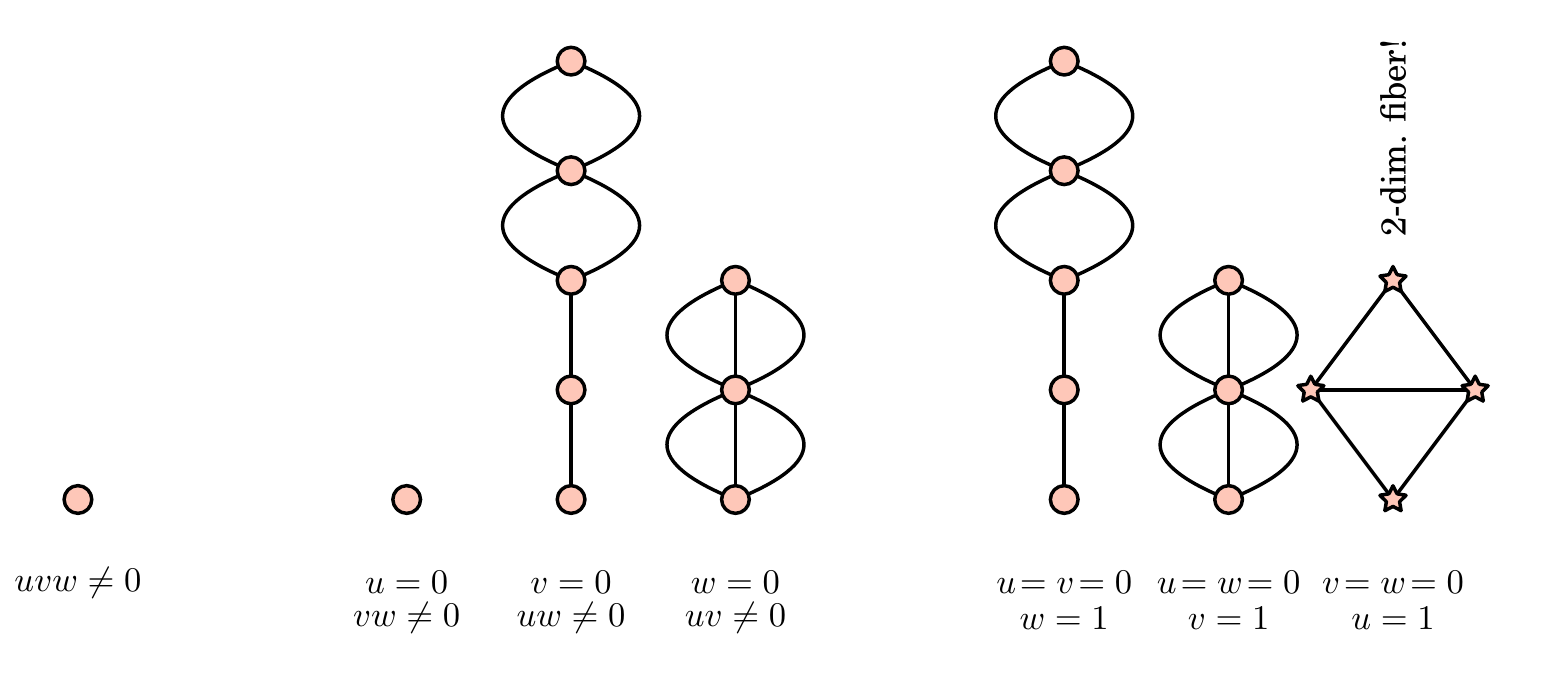}
  \caption{The fiber-divisor-graph $F(O(\sigma), -K)$ of the
    $2IV^*$-fibration over the $7$ torus-orbits $O(\sigma)$, $\sigma
    \in \Sigma$, in the base $\CP^2=\CP_\Sigma$.}
  \label{fig:fdgNonFlat}
\end{figure}
This asymmetry between the two $IV^*$ discriminant components is also
visible from the fiber-divisor-graph, see
\autoref{fig:fdgNonFlat}. The Kodaira diagram for the $IV^*$
degenerate fiber is the extended $\tilde{E}_6$ Dynkin diagram, which
is folded in two different ways into the fiber-divisor graph over the
$v=0$ and $w=0$ component of the discriminant. Over the intersection
point $v=w=0$ of the two $IV^*$ components of the discriminant the
elliptic fiber becomes complex two-dimensional and consists of $4$
irreducible components.


\section{Classification of Gauge Groups}

\subsection{Kodaira Fibers}

Having understood the structure of the elliptic fibration in terms of
the defining polytope, we can now compute the gauge groups arising
from each of the \comma{102581} flat toric elliptic fibrations.

Starting with a fibered reflexive lattice polytope $F\subset P$ with
$\dim F=2$, $\dim P = 4$, that admits a flat fibration over $\CP^2$ by
a toric morphism $\varphi$, we
\begin{enumerate}
\item Construct the face fan of $P$,
\item Subdivide the face fan to become a fibration over $\CP^2$,
\item Pick all integral points of $p\in P$ such that $\varphi(p)$ is
  zero or contained in a one-dimensional cone of the base $\CP^2$. In
  other words, all points that do not map into the interior of a
  two-cone of $\CP^2$.
\item Refine the fan further, using these additional rays.
\end{enumerate}
Proceeding this way, we can always resolve the ambient toric variety
far enough such that there are homogeneous coordinates that map to the
homogeneous coordinates of the base $\CP^2$, unlike the issue we
encountered in eq.~\eqref{eq:toricmorphismroot}. It is then
straightforward to compute the Weierstrass form of the hypersurface
equation and apply Tate's algorithm \autoref{tab:Tate}.
\begin{table}
  \centering
  \begin{tabular}{c|c}
    Fiber & \#
    \\ \hline
    $I_{2}$ & 53272  \\
    $I_{3}$ & 24303  \\
    $I_{4}$ & 42210  \\
    $I_{5}$ & 18981  \\
    $I_{6}$ & 28782  \\
    $I_{7}$ & 12884  \\
    $I_{8}$ & 15883  \\
    $I_{9}$ & 7424  \\
    $I_{10}$ & 7551  \\
    $I_{11}$ & 3325  \\
    $I_{12}$ & 3629  \\
    $I_{13}$ & 1288  \\
  \end{tabular}
  \hfill
  \begin{tabular}{c|c}
    Fiber & \#
    \\ \hline
    $I_{14}$ & 1364  \\
    $I_{15}$ & 519  \\
    $I_{16}$ & 537  \\
    $I_{17}$ & 150  \\
    $I_{18}$ & 207  \\
    $I_{19}$ & 37  \\
    $I_{20}$ & 71  \\
    $I_{21}$ & 15  \\
    $I_{22}$ & 17  \\
    $I_{23}$ & 1  \\
    $I_{24}$ & 11  \\
    $I_{27}$ & 1  \\
  \end{tabular}
  \hfill
  \begin{tabular}{c|c}
    Fiber & \#
    \\ \hline
    $I^*_{0}$ & 3803  \\
    $I^*_{1}$ & 2333  \\
    $I^*_{2}$ & 1971  \\
    $I^*_{3}$ & 1250  \\
    $I^*_{4}$ & 1030  \\
    $I^*_{5}$ & 596  \\
    $I^*_{6}$ & 477  \\
    $I^*_{7}$ & 249  \\
    $I^*_{8}$ & 204  \\
    $I^*_{9}$ & 92  \\
    $I^*_{10}$ & 77  \\
    $I^*_{11}$ & 31  \\
  \end{tabular}
  \hfill
  \begin{tabular}{c|c}
    Fiber & \#
    \\ \hline
    $I^*_{13}$ & 11  \\
    $I^*_{14}$ & 13  \\
    $I^*_{15}$ & 3  \\
    $I^*_{16}$ & 6  \\
    $I^*_{18}$ & 2  \\
    $I^*_{20}$ & 1  \\
    \multicolumn{2}{c}{} \\
    Fiber & \#
    \\ \hline
    $II^*$ & 100 \\ 
    $III^*$ & 429 \\
    $IV^*$ & 654 \\
  \end{tabular}
  \caption{Kodaira fibers in toric elliptic fibrations and their prevalence.}
  \label{tab:count}
\end{table}

\subsection{Transitions Among Vacua}

\begin{table}
  \centering
  \small
  \begin{tabular}{c|l}
    $h^{21}-h^{11}$ & 
    $(h^{11}, h^{21})$ \\
    \hline
    270 & (2, 272) \\
    228 & (3, 231) \\
    204 & (4, 208) \\
    192 & (3, 195) \\
    190 & (4, 194) \\
    184 & (5, 189) \\
    174 & (6, 180) \\
    168 & (5, 173) \\
    165 & (6, 171) \\
    162 & (3, 165) \\
    160 & (7, 167) \\
    158 & (4, 162) \\
    156 & (5, 161) \\
    153 & (8, 161) \\
    150 & (4, 154), (6, 156) \\
    147 & (6, 153), (7, 154) \\
    144 & (7, 151), (9, 153) \\
    142 & (2, 144), (8, 150) \\
    140 & (4, 144), (7, 147) \\
  \end{tabular}
  \hspace{1cm}
  \begin{tabular}{c|l}
    $h^{21}-h^{11}$ & 
    $(h^{11}, h^{21})$ \\
    \hline
    138 & (3, 141), (5, 143), \\
        & (8, 146), (10, 148) \\
    136 & (6, 142) \\
    133 & (7, 140) \\
    132 & (4, 136), (13, 145), (15, 147) \\
    130 & (2, 132), (7, 137), (8, 138) \\
    128 & (5, 133), (8, 136) \\
    126 & (2, 128), (6, 132), (18, 144) \\
    124 & (5, 129), (11, 135) \\
    122 & (4, 126), (6, 128) \\
    120 & (3, 123), (5, 125), (7, 127), \\
        & (9, 129), (10, 130), (14, 134), \\
        & (23, 143) \\
    117 & (4, 121), (7, 124), (8, 125) \\
    116 & (3, 119) \\
    114 & (5, 119), (6, 120), (8, 122), \\
        & (9, 123) \\
    112 & (4, 116), (5, 117), (7, 119), \\
        & (9, 121), (11, 123) \\
  \end{tabular}
  \\[5mm]
  \begin{tabular}{c|l}
    $h^{21}-h^{11}$ & 
    $(h^{11}, h^{21})$ \\
    \hline
    -47 & (62, 15) \\
    -48 & (54, 6), (55, 7), (56, 8), \\
        & (57, 9), (58, 10), (59, 11), \\
        & (60, 12), (61, 13), (62, 14), \\
        & (63, 15), (64, 16), (65, 17) \\
    -50 & (63, 13), (65, 15), (66, 16) \\
    -51 & (60, 9), (61, 10), (62, 11), \\
        & (63, 12), (64, 13), (66, 15) \\
    -52 & (64, 12), (67, 15) \\
    -54 & (60, 6), (61, 7), (62, 8), \\
        & (63, 9), (64, 10), (65, 11), \\
        & (66, 12), (67, 13), (68, 14) \\
    -56 & (67, 11), (68, 12), (69, 13), \\
        & (71, 15) \\
    -57 & (66, 9), (67, 10), (68, 11), \\
        & (70, 13) \\
    -58 & (68, 10), (70, 12) \\
  \end{tabular}
  \hspace{10mm}
  \begin{tabular}{c|l}
    $h^{21}-h^{11}$ & 
    $(h^{11}, h^{21})$ \\
    \hline
    -60 & (67, 7), (68, 8), (69, 9),  \\
        & (70, 10), (71, 11), (72, 12), \\
        & (73, 13) \\
    -63 & (71, 8), (72, 9), (73, 10), \\
        & (74, 11) \\
    -66 & (72, 6), (73, 7), (74, 8), \\
        & (75, 9), (76, 10), (77, 11),\\
        & (78, 12) \\
    -68 & (78, 10) \\
    -69 & (78, 9), (79, 10) \\
    -72 & (79, 7), (80, 8) \\
    -75 & (84, 9) \\
    -78 & (85, 7), (86, 8), (88, 10) \\
    -84 & (90, 6), (91, 7) \\
    -90 & (97, 7) \\
    -96 & (101, 5) \\
    -108 & (112, 4) \\
  \end{tabular}
  \hspace{5mm}
  \caption{Hodge numbers $(h^{11}, h^{21})$ of flat toric elliptic fibrations over
    $\CP^2$ for  $h^{21}-h^{11}\geq 112$ and $\leq -47$.}
  \label{tab:hodgenos}
\end{table}
The Hodge numbers of the flat toric elliptic fibrations over $\CP^2$
are shown in Figures~\ref{fig:bigplot} and~\ref{fig:zoomplot}. Let us
quickly note some of the salient features. The largest height
$h^{11}+h^{21} = 2+272$ is attained by a well-known elliptic
fibration, the resolution of the weighted projective space
$\CP^4[1,1,1,6,9]$~\cite{Vafa:1996xn, Candelas:1996ht,
  Morrison:1996pp, Morrison:1996na, Klemm:1996hh}. The next largest
Hodge numbers fall into a sequence $h^{11} = 2+k$, $h^{21} =
272-29k$~\cite{Candelas:1996su}. The factor of 29 is of course the
same as in the 6-d anomaly cancellation condition
\begin{equation}
  n_H - n_V = 273 - 29 n_T.
\end{equation}
as any transition between vacua has to preserve the anomaly. However,
increasing $n_T$ means that the base is blown up, so these sequences
are not visible when one restricts to a fixed base. If there are any
vacuum transition left after imposing the base, it should hold
$h^{21}-h^{11}$ constant. In \autoref{tab:hodgenos}, we list the Hodge
numbers for the left and right-most cases of the plot
\autoref{fig:bigplot}. While there does not seem to be any pattern to
the Hodge numbers with large and positive differences, the Hodge pairs
for large negative difference seem to come in sequences $(h^{11}+k,
h^{21}-k)$ for consecutive integers $k$. These are visible as vertical
lines in \autoref{fig:zoomplot}. Clearly, this is the usual Higgs
mechanism giving mass to both a vector and a hyper
multiplet. Moreover, large gauge groups are only on the side of large
numbers of vector multiplets, $h^{11}\gg 0$. This is nicely
illustrated by the fact that the vertical lines in
\autoref{fig:zoomplot} are only visible on the right-hand side of the
plot.

A mysterious pattern of the Hodge pairs with $h^{11}\gg h^{21}$ is
that they fall into linear sequences $(h^{11}+k, h^{21}-11k)$. For
example, the sequence starting with the manifold at the extreme right
is
\begin{equation}
  \big(
  (112, 4), (101, 5), (90, 6), (79, 7), (68, 8), 
  (57, 9), (46, 10), (35, 11), (24, 12)
  \big)
\end{equation}
and all are realized as Hodge numbers of elliptic fibrations. This
also holds true for the next right-most manifolds, for example
\begin{equation}
  \begin{gathered}
    \big(
    (97, 7), (86, 8), (75, 9), (64, 10), 
    (53, 11), (42, 12), (31, 13), (20, 14)
    \big)
    \\
    \big(
    (91, 7), (80, 8), (69, 9), (58, 10), 
    (47, 11), (36, 12), (25, 13)
    \big)
    \\
    \big(
    (85, 7), (74, 8), (63, 9), (52, 10), (41, 11), (30, 12), (19, 13)
    \big)
  \end{gathered}
\end{equation}

\subsection{SU(27) and Anomaly Cancellation}

The right-most Hodge pair $(h^{11},h^{21}) = (112,4)$ is realized by a
single fibered polytope and is in many ways analogous to our
simple-most example in \autoref{sec:example}. The lattice polytope is
spanned by the vertices
\begin{equation}
  P = \conv\big\{
  (-7, -7, -6, -9), (0, 1, 0, 0), (1, 0, 0, 0), 
  (2, 2, 3, 0), (2, 2, 3, 9)
  \big\}
\end{equation}
and contains the fiber sub-lattice polytope
\begin{equation}
  \conv\big\{
  (-2, -2, -2, -3), (1, 1, 1, 0), (1, 1, 1, 3)
  \big\} \subset P
\end{equation}
The polytope $P$ contains 67 integral points:
\begin{itemize}
\item The origin,
\item One point over the $\vec{u}$ and one over $\vec{v}$ in the base
  $\CP^2$ fan,
\item 55 points over $\vec{w}$, all being contained in a single
  two-face $F$,
\item and 10 points in the fiber sub-polytope (one of which is the
  origin).
\end{itemize}
\begin{figure}
  \centering
  \includegraphics{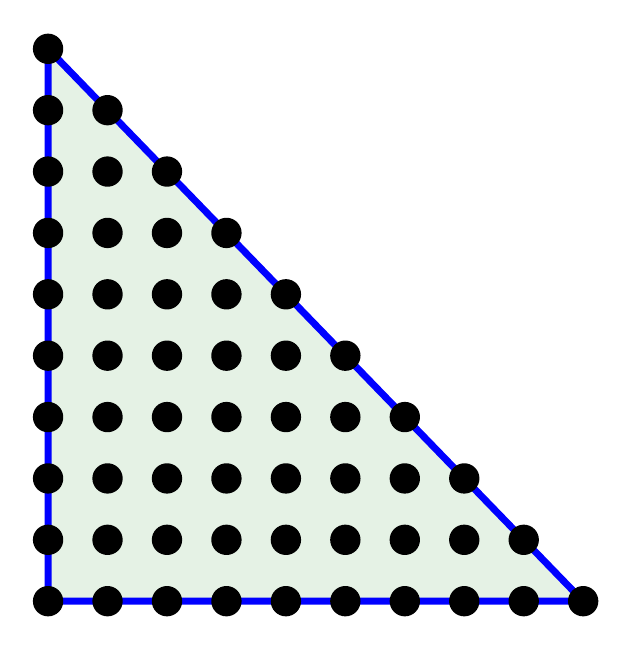}
  \caption{The two-face $F\subset P$ over $\vec{w}$ giving rise to the
    $I_{27}$ discriminant component.}
  \label{fig:SU27twoface}
\end{figure}
So the fiber is a cubic in $\CP^2/(\Z_3\times\Z_3)$, the mirror of a
cubic in $\CP^2$. There are no issues with remaining singularities;
One can completely resolve the fan into 243 smooth 4-cones while
preserving the fibration structure, and the subdivided fan is a flat
toric fibration over $\CP^2$ with respect to the lattice map
\begin{equation}
  \varphi = 
  \begin{pmatrix}
    0 & 1 & -1 & 0 \\
    1 & 0 & -1 & 0
  \end{pmatrix}
  .
\end{equation}
As always, we label the rays of the fan of $\CP^2$ as in
\autoref{fig:P2}. By the arguments above, the toric divisors $u=0$
and $v=0$ do not support a component of the discriminant, only $w=0$
does. The fiber-divisor-graph over $w=0$ is the $\hat{A}_{26}$
extended Dynkin diagram, that is, $27$ nodes in a circle. Just as in
\autoref{sec:example}, the extended Dynkin diagram can be seen as the
boundary of the two-face of the polytope that sits over $\vec{w}$ in
the base fan, see \autoref{fig:SU27twoface}. Hence the discriminant
component is a split $I_{27}$, leading to a $SU(27)$ gauge
theory. Alternatively, one can compute the Weierstrass form of the
hypersurface equation and arrive at the same conclusion.

However, in a theory without tensor multiplets the $SU(N)$ gauge group
is restricted to $N\leq 24$ by anomaly
cancellation~\cite{Morrison:2011mb, Kumar:2010am}. The resolution to
this puzzle is that there are extra tensor multiplets coming from a
type of codimension-two degeneration that is very generic in toric
elliptic fibrations but we have not discussed so far in this paper. In
the example under consideration, the toric fiber over $\cone{w}$
consists of 55 irreducible components, corresponding to the 55
integral points in $F$. The restriction of the anticanonical divisor
class is trivial on the 28 internal points, and non-trivial on the 27
points on the boundary of $F$. As we already mentioned before, this is
why the hypersurface equation will generically be 27 $\CP^1$ in
complex 2-dimensional toric fiber. The anticanonical divisor class
being trivial on a given irreducible toric fiber component means that
the hypersurface equation is constant, because that is the only
section of a trivial line bundle. But the constant may vary as one
moves the fiber around. In particular, the discriminant locus $w=0$ is
a $\CP^1$, so said constant varies in a one-parameter family. Unless
this constant along the fiber does not vary at all as one moves in the
base direction, there will be certain points of codimension two in the
base where the constant vanishes. This means that the fiber of the
Calabi-Yau threefold over this point includes a whole toric
surface. So while the toric fibration was flat, the elliptic fibration
is not\footnote{In other words, we classified flat toric elliptic
  fibrations in this paper and not toric flat elliptic fibrations.}
because the hypersurface equation identically vanishes over some
codimension-two point in the base.

Explicitly, let us divide the set of $66$ homogeneous coordinates into
\begin{itemize}
\item $u$ and $v$, the (unique) homogeneous coordinates whose rays map
  to the base $\vec{v}$ and $\vec{u}$.
\item $e_0$, $\dots$, $e_8$ the homogeneous coordinates on the fiber
  $\CP^2/(\Z_3\times\Z_3)$,
\item $f_0$, $\dots$, $f_{26}$ the homogeneous coordinates
  corresponding to the points on the boundary of the two-face $F$,
\item and $i_0$, $\dots$, $i_{27}$ the homogeneous coordinates
  corresponding to the points in the relative interior of the two-face
  $F$.
\end{itemize}
The hypersurface equation contains 13 coefficients $a_0$, $\dots$,
$a_{12}$. To set notation and for future reference, the hypersurface
equation reads in the $e_\bullet=f_\bullet=1$ patch:
\begin{equation}
  \begin{split}
    p =\;&
    a_{0} i_{0} i_{1}^{2} i_{2} i_{3}^{3} i_{4}^{2} i_{5} i_{6}^{4} i_{7}^{3} i_{8}^{2} i_{9} i_{10}^{5} i_{11}^{4} i_{12}^{3} i_{13}^{2} i_{14} i_{15}^{6} i_{16}^{5} i_{17}^{4} i_{18}^{3} i_{19}^{2} i_{20} i_{21}^{7} i_{22}^{6} i_{23}^{5} i_{24}^{4} i_{25}^{3} i_{26}^{2} i_{27} + 
    \\&
    a_{1} i_{0}^{7} i_{1}^{6} i_{2}^{6} i_{3}^{5} i_{4}^{5} i_{5}^{5} i_{6}^{4} i_{7}^{4} i_{8}^{4} i_{9}^{4} i_{10}^{3} i_{11}^{3} i_{12}^{3} i_{13}^{3} i_{14}^{3} i_{15}^{2} i_{16}^{2} i_{17}^{2} i_{18}^{2} i_{19}^{2} i_{20}^{2} i_{21} i_{22} i_{23} i_{24} i_{25} i_{26} i_{27} + 
    \\&
    a_{2} i_{0}^{3} i_{1}^{3} i_{2}^{3} i_{3}^{3} i_{4}^{3} i_{5}^{3} i_{6}^{3} i_{7}^{3} i_{8}^{3} i_{9}^{3} i_{10}^{3} i_{11}^{3} i_{12}^{3} i_{13}^{3} i_{14}^{3} i_{15}^{3} i_{16}^{3} i_{17}^{3} i_{18}^{3} i_{19}^{3} i_{20}^{3} i_{21}^{3} i_{22}^{3} i_{23}^{3} i_{24}^{3} i_{25}^{3} i_{26}^{3} i_{27}^{3} + 
    \\&
    a_{3} v i_{0}^{2} i_{1}^{2} i_{2}^{2} i_{3}^{2} i_{4}^{2} i_{5}^{2} i_{6}^{2} i_{7}^{2} i_{8}^{2} i_{9}^{2} i_{10}^{2} i_{11}^{2} i_{12}^{2} i_{13}^{2} i_{14}^{2} i_{15}^{2} i_{16}^{2} i_{17}^{2} i_{18}^{2} i_{19}^{2} i_{20}^{2} i_{21}^{2} i_{22}^{2} i_{23}^{2} i_{24}^{2} i_{25}^{2} i_{26}^{2} i_{27}^{2} + 
    \\&
    a_{4} v^{2} i_{0} i_{1} i_{2} i_{3} i_{4} i_{5} i_{6} i_{7} i_{8} i_{9} i_{10} i_{11} i_{12} i_{13} i_{14} i_{15} i_{16} i_{17} i_{18} i_{19} i_{20} i_{21} i_{22} i_{23} i_{24} i_{25} i_{26} i_{27} + 
    \\&
    a_{5} v^{3} + 
    \\&
    a_{6} u i_{0}^{2} i_{1}^{2} i_{2}^{2} i_{3}^{2} i_{4}^{2} i_{5}^{2} i_{6}^{2} i_{7}^{2} i_{8}^{2} i_{9}^{2} i_{10}^{2} i_{11}^{2} i_{12}^{2} i_{13}^{2} i_{14}^{2} i_{15}^{2} i_{16}^{2} i_{17}^{2} i_{18}^{2} i_{19}^{2} i_{20}^{2} i_{21}^{2} i_{22}^{2} i_{23}^{2} i_{24}^{2} i_{25}^{2} i_{26}^{2} i_{27}^{2} + 
    \\&
    a_{7} u v i_{0} i_{1} i_{2} i_{3} i_{4} i_{5} i_{6} i_{7} i_{8} i_{9} i_{10} i_{11} i_{12} i_{13} i_{14} i_{15} i_{16} i_{17} i_{18} i_{19} i_{20} i_{21} i_{22} i_{23} i_{24} i_{25} i_{26} i_{27} + 
    \\&
    a_{8} u v^{2} + 
    \\&
    a_{9} u^{2} i_{0} i_{1} i_{2} i_{3} i_{4} i_{5} i_{6} i_{7} i_{8} i_{9} i_{10} i_{11} i_{12} i_{13} i_{14} i_{15} i_{16} i_{17} i_{18} i_{19} i_{20} i_{21} i_{22} i_{23} i_{24} i_{25} i_{26} i_{27} + 
    \\&
    a_{10} u^{2} v + a_{11} u^{3} + 
    \\&
    a_{12} i_{0} i_{1} i_{2}^{2} i_{3} i_{4}^{2} i_{5}^{3} i_{6} i_{7}^{2} i_{8}^{3} i_{9}^{4} i_{10} i_{11}^{2} i_{12}^{3} i_{13}^{4} i_{14}^{5} i_{15} i_{16}^{2} i_{17}^{3} i_{18}^{4} i_{19}^{5} i_{20}^{6} i_{21} i_{22}^{2} i_{23}^{3} i_{24}^{4} i_{25}^{5} i_{26}^{6} i_{27}^{7}
  \end{split}
\end{equation}
and the toric morphism is
\begin{equation}
  \Sigma_P \to \CP^2,
  \quad
  [u:v:e_\bullet:f_\bullet:i_\bullet]
  \mapsto
  \Big[u: v: 
  \prod e_\bullet \prod f_\bullet \prod i_\bullet \Big]
\end{equation}
The complex 3-dimensional toric divisor $i_j=0$ maps onto the base
toric divisor $w=0$, so its fibers are 2-dimensional. For a fixed base
point $[u:v:0]$, these are the 28 irreducible components of the toric
fiber that correspond to the interior points of the two-face $F\subset
P$. The pull-back of the anticanonical class on these toric fiber
components is trivial, so the section is constant for fixed $u$,
$v$. To determine the constant, we evaluate\footnote{Of course these
  are sections of bundles, so strictly speaking it does not make sense
  to ``evaluate'' them. What is well-defined, however, is to test
  whether they are zero or not.} the hypersurface equation at a
generic point, that is, a point in the maximal torus orbit of the
toric fiber component. In other words, set
\begin{equation}
  i_j = 0
  ,\quad
  i_k = 1~ \forall k\not=j
  ,\quad
  e_\bullet = f_\bullet = 1
  .
\end{equation}
Independent of the which of the $28$ $i_j$ we set to zero, the
hypersurface equation becomes
\begin{equation}
  p\big(i_j=0,~ i_k = 1~ \forall k\not=j,~ 
  e_\bullet = f_\bullet = 1\big)
  = a_{5} v^{3} + a_{8} u v^{2} + a_{10} u^{2} v + a_{11} u^{3}
  .
\end{equation}
So the constant vanishes at the three solutions of the above cubic. To
summarize, there is an $I_{27}$ Kodara fiber over $\{w=0\}\simeq
\CP^1$. Over three points along this discriminant locus, the fiber
jumps in dimension and becomes a reducible 2-dimensional toric variety
with $28$ irreducible components.

\appendix

\section{Weierstrass Forms}
\label{sec:Weierstrass}

\begin{figure}[htbp]
  \centering
  \includegraphics{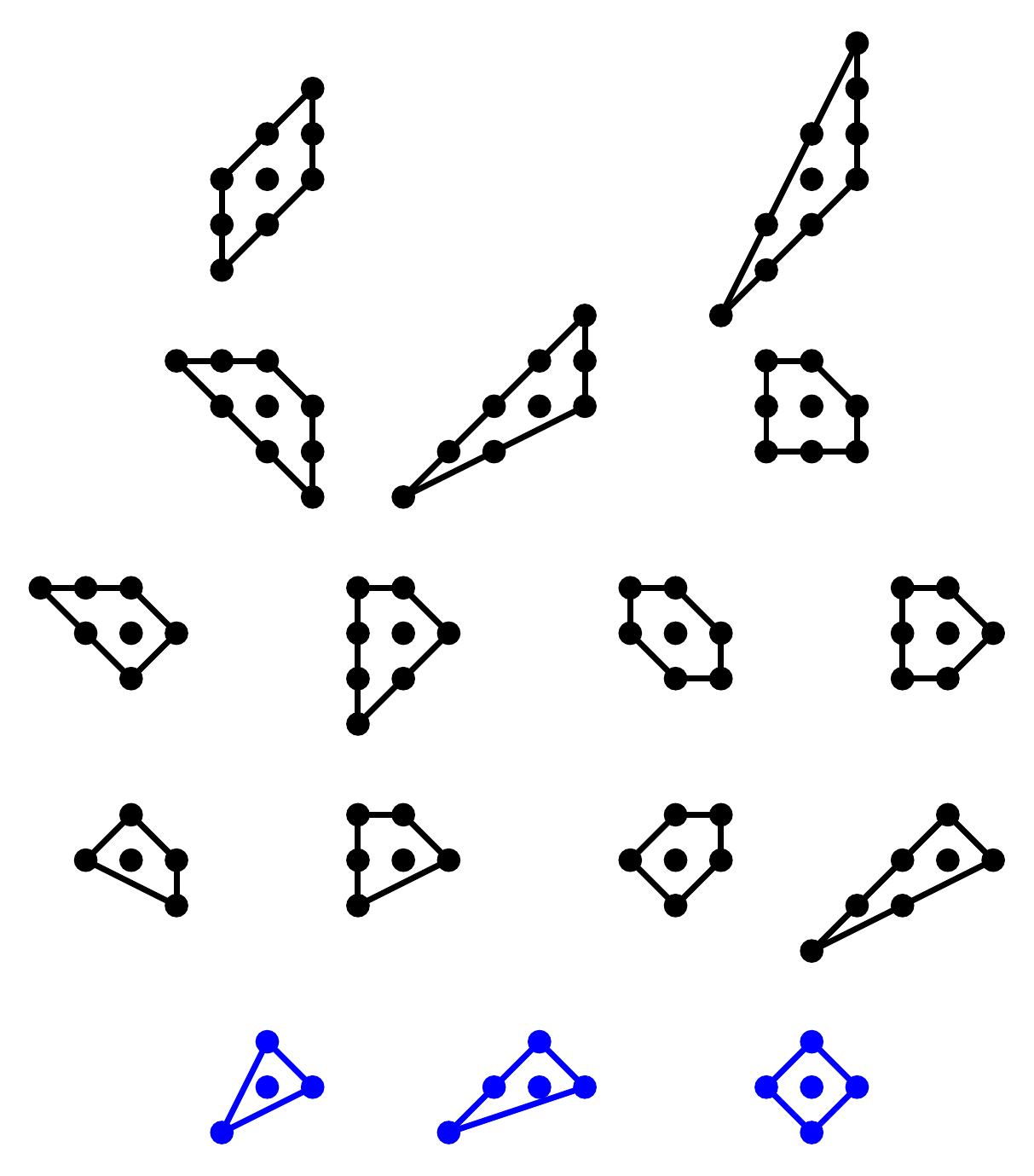}
  \caption{The $16$ reflexive lattice polygons. The $3$ blue polygons
    at the bottom row are the ones that do not contain a smaller
    reflexive polygon.}
  \label{fig:reflexive2d}
\end{figure}

Consider a fibration of lattice polytopes $F\hookrightarrow P$. If the
lattice polytope $P$ is reflexive, then the lattice sub-polytope $F$
is reflexive, too. For the purposes of this paper, the fiber polytope
will always be $2$-dimensional, that is, one of the $16$ reflexive
polygons shown in \autoref{fig:reflexive2d}. 
Each lattice polygon defines a face fan and therefore a 2-dimensional
compact toric variety. In 2 dimensions, there is a unique maximal
cepant desingularization by subdividing the fan such that all lattice
points on the boundary of the polygon span a ray of the fan. A generic
section of the anticanonical divisor then defines a smooth Calabi-Yau
one-fold (that is, a real $2$-torus), irregardless of whether or not
one resolves the point-singularities in the ambient toric variety. A
smooth $2$-torus can be written as a cubic in $\CP^2$, where the cubic
can be taken to be in Weierstrass from $y^2=x^3+ax+b$.

In order to identify the discriminant locus of the toric elliptic
firations, we need to be able to explictly write the Calabi-Yau
hypersurface equation in Weierstrass from. First, however, note that
we do not need to give equations for the Weierstrass from for all $16$
reflexive lattice polygons. Since the monomials of the anticanonical
hypersurface are the integral points of the dual lattice polygon, we
only need to find the transformation to Weierstrass form for the
minimal polygons with respect to inclusion. Any strictly larger
polygon has a strictly smaller dual polygon, so its anticanonical
hypersurface equation is just a specialization where some coefficients
are set to zero. In fact, there are $3$ minimal reflexive lattice
polytopes, which are shown in blue in \autoref{fig:reflexive2d}. The
corresponding toric varieties are $\CP^2$, the weighted projective
plane $\CP^2[1,1,2]$, and $\CP^1\times\CP^1$. In the remainder of this
appendix, we will discuss these three cases:
\begin{itemize}
\item Transforming a cubic in $\CP^2$ into Weierstrass is well-known,
  and many computer algebra systems provide an
  implementation.
\item An anticanonical hypersurface in $\CP^1\times\CP^1$ is a
  biquadric eq.~\eqref{eq:biquadric}. The Weierstrass form of the
  elliptic curve embedded as a hypersurface in $\CP^2$ is given in
  eq.~\eqref{eq:WeierstrassBiquadric}~\cite{pre05771645}.
\item The remaining case of an anticanonical hypersurface in weighted
  projective space $\CP^2[1,1,2]$ will be treated shortly.
\end{itemize}
Counting only the degrees of the homogeneous coordinates in the fiber
fan, the Newton polytope of the hypersurface equation of a toric
elliptic fibered Calabi-Yau is always a sub-polytope of the dual
polytope of $\CP^2$ (27 sub-polytopes), the dual polytope of
$\CP^1\times\CP^1$ (20 sub-polytopes), or of the dual polytope of
$\CP^2[1,1,2]$ (28 sub-polytopes). By embedding the Newton polytope of
the hypersurface equation we can then easily compute the Weierstrass
form of the hypersurface using the same coordinate transformations as
the containing (maximal) reflexive lattice polytope.

It remains to find the Weierstrass cubic representation of an
anticanonical hypersurface in $\CP^2[1,1,2]$. 
\begin{figure}
  \centering
  \includegraphics{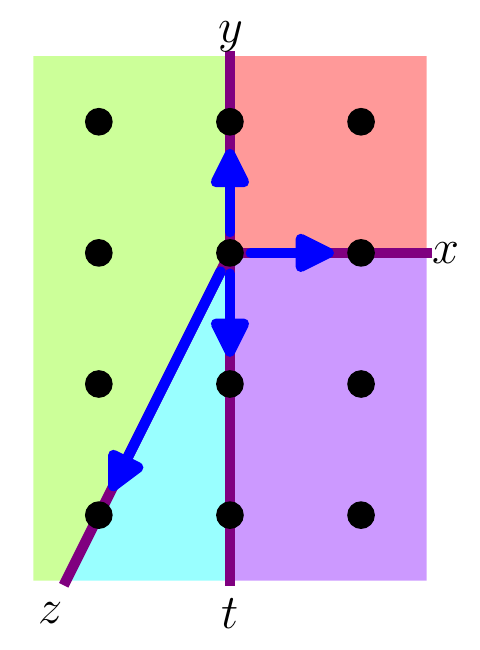}$~$\\
  $\vcenter{\hbox{\Huge $\downarrow$}}\;\varphi$\\
  \vspace{-10mm}
  \includegraphics{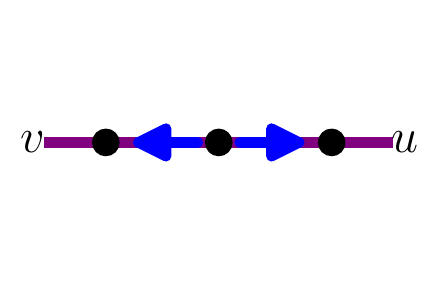}
  \vspace{-7mm}
  \caption{Toric fibration of the resolved weighted projective space
    $\widehat{\CP^2[1,1,2]}$ over $\CP^1$.}
  \label{fig:P112fib}
\end{figure}
Note that there is a single fibration of the resolved $\CP^2[1,1,2]$
shown in \autoref{fig:P112fib}, which suggests to take first the
discriminant along the fiber directions as in the $\CP^1\times\CP^1$
case. The $9$ sections of the anticanonical bundle are
\begin{equation}
  H^0\left( \widehat{\CP^2[1,1,2]}, -K \right) =
  \Span\big\{
  y^2, 
  y z^2 t, 
  x y z t, 
  x^2 y t, 
  z^4 t^2, 
  x z^3 t^2, 
  x^2 z^2 t^2, 
  x^3 z t^2, 
  x^4 t^2
  \big\}
\end{equation}
For convenience, let us switch to inhomogeneous coordinates where
$z=t=1$, then the hypersurface equation for an elliptic curve reads
\begin{equation}
  C(x,y) = 
  \alpha_{40} x^4 + 
  \alpha_{30} x^3 + 
  \alpha_{21} x^2 y + 
  \alpha_{20} x^2 + 
  \alpha_{11} x y + 
  \alpha_{02} y^2 + 
  \alpha_{10} x + 
  \alpha_{01} y + 
  \alpha_{00}
\end{equation}
It is quadratic in $y$ with the ordinary quadratic discriminant
\begin{equation}
  \beta_4 x^4 + \beta_3 x^3 + \beta_2 x^2 + \beta_1 x + \beta_0 =
  \left(
    \sum \alpha_{i1} x^i
  \right)^2
  - 4
  \left(
    \sum \alpha_{i2} x^i
  \right)
  \left(
    \sum \alpha_{i0} x^i
  \right)
  .
\end{equation}
Again, the quadratic discriminant is a plane quartic as in
eq.~\eqref{eq:biquadric}. The coefficients $a$, $b$ of the Weierstrass
form $y^2=x^3+ax+b$ are then again given by the quadratic and cubic
projective $GL(2,\C)$-invariants,
\begin{equation}
  \begin{split}
    a =& - \tfrac{1}{4} \big(
    \beta_0 \beta_4 + 3 \beta_2^2 - 4 \beta_1 \beta_3
    \big) \\
    b =& - \tfrac{1}{4} \big(
    \beta_0 \beta_3^2 +\beta_1^2 \beta_4 -\beta_0 \beta_2 \beta_4 
    -2 \beta_1 \beta_2 \beta_3 +\beta_2^3
    \big)
    .
  \end{split}
\end{equation}


\bibliographystyle{utcaps} 
\renewcommand{\refname}{Bibliography}
\addcontentsline{toc}{section}{Bibliography} 
\bibliography{Main}

\providecommand{\href}[2]{#2}\begingroup\raggedright\begin{thebibliography}{10}

\bibitem{MR0393039}
J.~Tate, ``Algorithm for determining the type of a singular fiber in an
  elliptic pencil,'' in {\em Modular functions of one variable, {IV} ({P}roc.
  {I}nternat. {S}ummer {S}chool, {U}niv. {A}ntwerp, {A}ntwerp, 1972)},
  pp.~33--52. Lecture Notes in Math., Vol. 476.
\newblock Springer, Berlin, 1975.

\bibitem{Vafa:1996xn}
C.~Vafa, ``{Evidence for F-Theory},'' {\em Nucl. Phys.} {\bf B469} (1996)
  403--418,
\href{http://arXiv.org/abs/hep-th/9602022}{{\tt hep-th/9602022}}.

\bibitem{Morrison:1996na}
D.~R. Morrison and C.~Vafa, ``{Compactifications of F theory on Calabi-Yau
  threefolds. 1},'' {\em Nucl.Phys.} {\bf B473} (1996) 74--92,
  \href{http://arXiv.org/abs/hep-th/9602114}{{\tt hep-th/9602114}}.

\bibitem{Morrison:1996pp}
D.~R. Morrison and C.~Vafa, ``{Compactifications of F theory on Calabi-Yau
  threefolds. 2.},'' {\em Nucl.Phys.} {\bf B476} (1996) 437--469,
  \href{http://arXiv.org/abs/hep-th/9603161}{{\tt hep-th/9603161}}.

\bibitem{Donagi:2009ra}
R.~Donagi and M.~Wijnholt, ``{Higgs Bundles and UV Completion in F-Theory},''
  \href{http://arXiv.org/abs/0904.1218}{{\tt 0904.1218}}.

\bibitem{Marsano:2011hv}
J.~Marsano and S.~Schafer-Nameki, ``{Yukawas, G-flux, and Spectral Covers from
  Resolved Calabi-Yau's},'' \href{http://arXiv.org/abs/1108.1794}{{\tt
  1108.1794}}. * Temporary entry *.

\bibitem{Katz:2011qp}
S.~Katz, D.~R. Morrison, S.~Schafer-Nameki, and J.~Sully, ``{Tate's algorithm
  and F-theory},'' {\em JHEP} {\bf 1108} (2011) 094,
  \href{http://arXiv.org/abs/1106.3854}{{\tt 1106.3854}}.

\bibitem{MR690264}
R.~Miranda, ``Smooth models for elliptic threefolds,'' in {\em The birational
  geometry of degenerations ({C}ambridge, {M}ass., 1981)}, vol.~29 of {\em
  Progr. Math.}, pp.~85--133.
\newblock Birkh\"auser Boston, Mass., 1983.

\bibitem{Esole:2011sm}
M.~Esole and S.-T. Yau, ``{Small resolutions of SU(5)-models in F-theory},''
\href{http://arXiv.org/abs/1107.0733}{{\tt 1107.0733}}.

\bibitem{Vafa:2005ui}
C.~Vafa, ``{The string landscape and the swampland},''
\href{http://arXiv.org/abs/hep-th/0509212}{{\tt hep-th/0509212}}.

\bibitem{Kumar:2009ac}
V.~Kumar, D.~R. Morrison, and W.~Taylor, ``{Mapping 6D N = 1 supergravities to
  F-theory},'' {\em JHEP} {\bf 02} (2010) 099,
\href{http://arXiv.org/abs/0911.3393}{{\tt 0911.3393}}.

\bibitem{Kumar:2010am}
V.~Kumar, D.~S. Park, and W.~Taylor, ``{6D supergravity without tensor
  multiplets},'' {\em JHEP} {\bf 1104} (2011) 080,
  \href{http://arXiv.org/abs/1011.0726}{{\tt 1011.0726}}.

\bibitem{Kumar:2010ru}
V.~Kumar, D.~R. Morrison, and W.~Taylor, ``{Global aspects of the space of 6D N
  = 1 supergravities},'' {\em JHEP} {\bf 1011} (2010) 118,
  \href{http://arXiv.org/abs/1008.1062}{{\tt 1008.1062}}.

\bibitem{Seiberg:2011dr}
N.~Seiberg and W.~Taylor, ``{Charge Lattices and Consistency of 6D
  Supergravity},'' {\em JHEP} {\bf 1106} (2011) 001,
  \href{http://arXiv.org/abs/1103.0019}{{\tt 1103.0019}}. * Temporary entry *.

\bibitem{2000math.....10082H}
Y.~{Hu}, C.-H. {Liu}, and S.-T. {Yau}, ``{Toric morphisms and fibrations of
  toric Calabi-Yau hypersurfaces},'' {\em ArXiv Mathematics e-prints} (Oct.,
  2000) \href{http://arXiv.org/abs/arXiv:math/0010082}{{\tt
  arXiv:math/0010082}}.

\bibitem{Cox:1993fz}
D.~A. Cox, ``The Homogeneous Coordinate Ring of a Toric Variety, Revised
  Version,''.

\bibitem{KreuzerSkarkeReflexive}
M.~Kreuzer and H.~Skarke, ``{Complete classification of reflexive polyhedra in
  four dimensions},'' {\em Adv. Theor. Math. Phys.} {\bf 4} (2002) 1209--1230,
\href{http://arXiv.org/abs/hep-th/0002240}{{\tt hep-th/0002240}}.

\bibitem{Braun:2011hd}
V.~Braun, ``{The 24-Cell and Calabi-Yau Threefolds with Hodge Numbers (1,1)},''
  \href{http://arXiv.org/abs/1102.4880}{{\tt 1102.4880}}.

\bibitem{Kreuzer:2002uu}
M.~Kreuzer and H.~Skarke, ``PALP: A Package for analyzing lattice polytopes
  with applications to toric geometry,'' {\em Comput. Phys. Commun.} {\bf 157}
  (2004) 87--106,
\href{http://arXiv.org/abs/math.na/0204356}{{\tt math.na/0204356}}.

\bibitem{Sage}
W.~A. Stein {\em et al.}, {\em {S}age {M}athematics {S}oftware ({V}ersion
  4.7)}.
\newblock The Sage Development Team, 2011.
\newblock {\tt http://www.sagemath.org}.

\bibitem{BraunHampton:polyhedra}
V.~Braun and M.~Hampton, {\em Polyhedra module of {S}age}.
\newblock The Sage Development Team, 2010.
\newblock {\tt http://sagemath.org/doc/reference/sage/geometry/polyhedra.html}.

\bibitem{KodairaII}
K.~Kodaira, ``On compact analytic surfaces II,'' {\em Annals of Math.} {\bf 77}
  (1963) 563--626.

\bibitem{KodairaIII}
K.~Kodaira, ``On compact analytic surfaces III,'' {\em Annals of Math.} {\bf
  78} (1963) 1--40.

\bibitem{dP9Z3Z3}
V.~Braun, B.~A. Ovrut, T.~Pantev, and R.~Reinbacher, ``Elliptic {{Calabi-Yau}}
  threefolds with {$\Z_3 \times \Z_3$} {W}ilson lines,'' {\em JHEP} {\bf 12}
  (2004) 062,
\href{http://arXiv.org/abs/hep-th/0410055}{{\tt hep-th/0410055}}.

\bibitem{Braun:2007vy}
V.~Braun, M.~Kreuzer, B.~A. Ovrut, and E.~Scheidegger, ``Worldsheet Instantons
  and Torsion Curves, Part B: Mirror Symmetry,'' {\em JHEP} {\bf 10} (2007)
  023,
\href{http://arXiv.org/abs/arXiv:0704.0449 [hep-th]}{{\tt arXiv:0704.0449
  [hep-th]}}.

\bibitem{Braun:2007xh}
V.~Braun, M.~Kreuzer, B.~A. Ovrut, and E.~Scheidegger, ``Worldsheet instantons
  and torsion curves. Part A: Direct computation,'' {\em JHEP} {\bf 10} (2007)
  022,
\href{http://arXiv.org/abs/hep-th/0703182}{{\tt hep-th/0703182}}.

\bibitem{Braun:2007tp}
V.~Braun, M.~Kreuzer, B.~A. Ovrut, and E.~Scheidegger, ``Worldsheet Instantons,
  Torsion Curves, and Non-Perturbative Superpotentials,'' {\em Phys. Lett.}
  {\bf B649} (2007) 334--341,
\href{http://arXiv.org/abs/hep-th/0703134}{{\tt hep-th/0703134}}.

\bibitem{Morrison:2011mb}
D.~R. Morrison and W.~Taylor, ``{Matter and singularities},''
\href{http://arXiv.org/abs/1106.3563}{{\tt 1106.3563}}.

\bibitem{pre05771645}
J.~J. Duistermaat, {\em {Discrete integrable systems. QRT maps and elliptic
  surfaces.}}
\newblock {Springer Monographs in Mathematics. Berlin: Springer. xxii, 627~p.},
  2010.

\bibitem{Bershadsky:1996nh}
M.~Bershadsky {\em et al.}, ``{Geometric singularities and enhanced gauge
  symmetries},'' {\em Nucl. Phys.} {\bf B481} (1996) 215--252,
\href{http://arXiv.org/abs/hep-th/9605200}{{\tt hep-th/9605200}}.

\bibitem{Grassi:2011hq}
A.~Grassi and D.~R. Morrison, ``{Anomalies and the Euler characteristic of
  elliptic Calabi-Yau threefolds},'' \href{http://arXiv.org/abs/1109.0042}{{\tt
  1109.0042}}.

\bibitem{Candelas:1996ht}
P.~Candelas, E.~Perevalov, and G.~Rajesh, ``{F theory duals of nonperturbative
  heterotic E(8) x E(8) vacua in six-dimensions},'' {\em Nucl.Phys.} {\bf B502}
  (1997) 613--628, \href{http://arXiv.org/abs/hep-th/9606133}{{\tt
  hep-th/9606133}}.

\bibitem{Candelas:1997pv}
P.~Candelas, E.~Perevalov, and G.~Rajesh, ``{Comments on A, B, C chains of
  heterotic and type II vacua},'' {\em Nucl. Phys.} {\bf B502} (1997) 594--612,
\href{http://arXiv.org/abs/hep-th/9703148}{{\tt hep-th/9703148}}.

\bibitem{Candelas:1997eh}
P.~Candelas, E.~Perevalov, and G.~Rajesh, ``{Toric geometry and enhanced gauge
  symmetry of F theory / heterotic vacua},'' {\em Nucl.Phys.} {\bf B507} (1997)
  445--474, \href{http://arXiv.org/abs/hep-th/9704097}{{\tt hep-th/9704097}}.

\bibitem{Candelas:1997pq}
P.~Candelas and H.~Skarke, ``{F theory, SO(32) and toric geometry},'' {\em
  Phys.Lett.} {\bf B413} (1997) 63--69,
  \href{http://arXiv.org/abs/hep-th/9706226}{{\tt hep-th/9706226}}.

\bibitem{Candelas:1997jz}
P.~Candelas, E.~Perevalov, and G.~Rajesh, ``{Matter from toric geometry},''
  {\em Nucl.Phys.} {\bf B519} (1998) 225--238,
  \href{http://arXiv.org/abs/hep-th/9707049}{{\tt hep-th/9707049}}.

\bibitem{Braun:2000hh}
V.~Braun, P.~Candelas, X.~De~La~Ossa, and A.~Grassi, ``{Toric Calabi-Yau
  fourfolds, duality between N = 1 theories and divisors that contribute to the
  superpotential},''
\href{http://arXiv.org/abs/hep-th/0001208}{{\tt hep-th/0001208}}.

\bibitem{Klemm:1996hh}
A.~Klemm, P.~Mayr, and C.~Vafa, ``{BPS states of exceptional non-critical
  strings},''
\href{http://arXiv.org/abs/hep-th/9607139}{{\tt hep-th/9607139}}.

\bibitem{Candelas:1996su}
P.~Candelas and A.~Font, ``{Duality between the webs of heterotic and type II
  vacua},'' {\em Nucl. Phys.} {\bf B511} (1998) 295--325,
\href{http://arXiv.org/abs/hep-th/9603170}{{\tt hep-th/9603170}}.

\end{thebibliography}\endgroup

\end{document}